\documentclass[11pt]{article}
\usepackage[a4paper, total={6in, 8in}]{geometry}
\usepackage{enumerate}
\usepackage{amsmath}
\usepackage{amssymb,latexsym}
\usepackage{amsthm}
\usepackage{color}
\usepackage{cancel}
\usepackage{graphicx}
\usepackage{mathtools}

\usepackage{hyperref}
\usepackage{comment}
\usepackage{amscd}
\usepackage{cite}
\usepackage{lineno}

\makeatletter
\usepackage{comment}
\let\wfs@comment@comment\comment
\let\comment\@undefined

\usepackage{changes}
\let\wfs@changes@comment\comment
\let\comment\@undefined

\newcommand\comment{%
    \ifthenelse{\equal{\@currenvir}{comment}}
    {\wfs@comment@comment}
    {\wfs@changes@comment}%
}

\definechangesauthor[name=Daniele, color=red]{DAN}
\definechangesauthor[name=Giovanni, color=blue]{GIO}
\definechangesauthor[name=Nico, color=green]{NICO}

\newtheorem*{theorem*}{Theorem}

\title{Homomorphic encryption schemes based on coding theory and polynomials}
\author{Giovanni Giuseppe Grimaldi\thanks{Department of Mathematics and Informatics, University of Perugia, Perugia, Italy.
Email address: {\tt{giovannigiuseppe.grimaldi@unipg.it}}}}
\date{}
\begin{document}

\maketitle

\begin{abstract}
Homomorphic encryption is a powerful cryptographic tool that enables secure computations on the private data. It 
evaluates any function for any operation securely on the 
encrypted data without knowing its corresponding plaintext.
For original data $p$, $c$ denotes the ciphertext of the original 
plaintext $p$, i.e. $c = Encrypt_k(p)$. This is crucial for any sensitive application running in the Cloud, because we must protect data privacy even in the case when the server has fallen victim to a cyber attack. The encryption scheme $Encrypt_k$ is said
to be homomorphic with respect to some set of operations $\mathcal{O}$, 
if for any operation $\circ \in \mathcal{O}$ one can compute $Encrypt_k(p_1 \circ p_2)$ from $Encrypt_k(p_1) \circ Encrypt_k(p_2)$. Those schemes come in three forms: somewhat, partially and fully homomorphic. In this survey, we present the state of art of the known homomorphic encryption schemes based on coding theory and polynomials.
\end{abstract}


\section{Introduction}
In algebra, a \textit{homomorphism} between two algebraic structures $(S_1, \star)$ and $(S_2, \circ)$ is defined as a map $F : S_1 \longrightarrow S_2$ such that
\[
F(a \star b)=F(a) \circ F(b),
\]
for all $a,b \in S_1$. In the context of cryptography, the term homomorphism refers to the process of encrypting messages. Let $\mathcal{M}$ be the message space and $\mathcal{F}$ be a set of operations. A cryptographic scheme $\mathcal{E}$ is said to be \textit{homomorphic with respect to $\mathcal{F}$} if, for every $f \in \mathcal{F}$ and $m_1,\ldots,m_k \in \mathcal{M}$, it is possible to determine $Encrypt(f(m_1,\ldots,m_k))$ from $Encrypt(m_1),\ldots,Encrypt(m_k),$ where $Encrypt$ denotes the encryption function.

Unlike traditional cryptography, a homomorphic cryptographic scheme allows a third party, such as a cloud service, to perform operations on encrypted data without decrypting it, and therefore is able to guarantee a level of data security and protection.

A homomorphic cryptographic scheme $\mathcal{E}$ has four algorithms: $Keygen_{\mathcal{E}}$, $Encrypt_{\mathcal{E}}$, $Decrypt_{\mathcal{E}}$, and $Eval_{\mathcal{E}}$. $Keygen_{\mathcal{E}}$, $Encrypt_{\mathcal{E}}$, and $Decrypt_{\mathcal{E}}$ are identical to those used in traditional cryptography, and thus we have:
\begin{itemize}
    \item $Keygen_{\mathcal{E}}$ takes a security parameter $\lambda$ as input and returns a secret key $\textbf{sk}$ and a public key $\textbf{pk}$ as output;
    \item $Encrypt_{\mathcal{E}}$ takes a message $m$ and the public key $\textbf{pk}$ as input and returns a ciphertext $c$ as output;
    \item $Decrypt_{\mathcal{E}}$ takes the ciphertext $c$ and the secret key $\textbf{sk}$ as input and returns the message $m$ as output.
\end{itemize}
The peculiarity of homomorphic cryptography lies in the presence of a fourth algorithm, $Eval_{\mathcal{E}}$, associated with a set $\mathcal{F}$ of operations. For every operation $f \in \mathcal{F}$ and for every ciphertext $c_1, \ldots , c_k$ with $c_i = Encrypt_{\mathcal{E}} (m_i)$, the algorithm $Eval_{\mathcal{E}}(f, c_1, \ldots , c_t)$ provides as output a ciphertext $c$ such that
\[Decrypt_{\mathcal{E}}(c) = f(m_1, \ldots , m_k).\]
As in classical cryptography, a homomorphic cryptographic scheme can be \textit{symmetric-key} or \textit{asymmetric-key}. Furthermore, we have that:
\begin{itemize}
    \item the scheme $\mathcal{E}$ is \textit{partially homomorphic} if $Eval_{\mathcal{E}}$ supports a single operation that can be performed an unlimited number of times on the data;
    \item the scheme $\mathcal{E}$ is \textit{somewhat homomorphic} if $Eval_{\mathcal{E}}$ supports an arbitrary number of operations that can be performed a limited number of times on the data;
    \item the scheme $\mathcal{E}$ is \textit{fully homomorphic} if $Eval_{\mathcal{E}}$ supports an arbitrary number of operations that can be performed an unlimited number of times on the data.
\end{itemize}

In addition to fully homomorphic encryption, the literature distinguishes \textit{leveled fully homomorphic encryption}. Unlike pure fully homomorphic schemes, leveled ones support the evaluation of circuits only up to a predefined depth fixed at key generation time. This limitation arises because each homomorphic operation (particularly multiplication) introduces noise into the ciphertext; once this noise exceeds a specific threshold, the underlying data can no longer be successfully decrypted.

A fundamental characteristic of homomorphic cryptography is that the encryption function is probabilistic. Therefore, multiple ciphertexts correspond to the same message.

In \cite{boot}, Gentry provided a \textit{bootstrapping} algorithm which decreases the errors produced in the ciphertexts after homomorphic operations. This algorithm can be applied to any homomorphic scheme to obtain a fully homomorphic scheme. However, bootstrapping is computationally complex and it is hard to implement in real applications.

\section{Preliminaries}
Let $\mathbb{F}$ be a field, an $[n,k]$-linear code $\mathcal{C}$ is a subspace of dimension $k$ of $\mathbb{F}^n$. The \textit{Hamming distance} of a vector $\textbf{v}=(v_1,\ldots,v_n)$ is $d_H(\textbf{v}) = |\{ i : v_i \ne 0 \}|$, and the \textit{minimum Hamming distance} of the code $\mathcal{C}$ is defined as
\[
d= \min \{ d_{H}(\textbf{v}) : \textbf{v} \in \mathcal{C} \setminus \{ \textbf{0} \} \}.
\]
In this case, we will say that $\mathcal{C}$ is an $[n,k,d]$-linear code. Any linear code $\mathcal{C}$ can be represented in two equivalent ways:
\begin{itemize}
    \item by a \textit{generator matrix} $\textbf{G} \in \mathbb{F}^{k \times n}$, whose rows constitute a basis of $\mathcal{C}$ and
    $$
    \mathcal{C}=\{ \textbf{xG} : \textbf{x} \in \mathbb{F}_q^k \};
    $$
    \item by a \textit{parity check matrix} $\textbf{H} \in \mathbb{F}^{(n-k) \times n}$, whose rows constitute a basis of the dual subspace of $\mathcal{C}$, and 
    $$
    \mathcal{C}=\{ \textbf{x} \in \mathbb{F}_q^n : \textbf{H}\textbf{x}^t=\textbf{0} \}.
    $$
\end{itemize}

In the following, we will denote the set $\{ 1,\ldots,n \}$ with $[n]$ and with
\[supp(\textbf{v})=\{ i : v_i \ne 0 \}\]
the \textit{support} of the vector $\textbf{v}=(v_1,\ldots,v_n)$. If $i \in supp(\textbf{v})$, then $i$ will be called a \textit{good location}. Otherwise, $i$ will be called a \textit{bad location}. A vector of a linear code $\mathcal{C}$ will be called an \textit{error-free codeword} or, simply, a \textit{codeword}. While, a vector of $\mathbb{F}^n \setminus \mathcal{C}$ will be called an \textit{erroneous codeword}. In particular, every erroneous codeword $\textbf{w}$ is of the form
\[
\textbf{w}=\textbf{v}+\textbf{e}
\]
where $\textbf{v} \in \mathcal{C}$, $\textbf{e} \in \mathbb{F}^n \setminus \{ \textbf{0} \}$ is called the \textit{error vector}.

With the symbol $\odot$ we will denote the component-wise product between two vectors, known as the \textit{Hadamard product}.

Now, let $\mathbb{F}_{q^m}$ an extension of degree $m$ of the finite field with $q$ elements $\mathbb{F}_q$. Let $\alpha$ be a root of an irreducible polynomial $f$ of degree $m$ over $\mathbb{F}_q$. Then, $\{1,\alpha,\alpha^2, \ldots, \alpha^{m-1}\}$ is an $\mathbb{F}_q$-basis of $\mathbb{F}_{q^m}$ and we may consider the map
$$
\textbf{vec}: x=\sum_{i=0}^{m-1}x_i \alpha^i \in \mathbb{F}_{q^m} \mapsto (x_0,x_1, \ldots, x_{m-1}) \in \mathbb{F}_{q}^m.
$$
Given $\textbf{v}=(v_1,v_2,\ldots,v_n) \in \mathbb{F}_{q^m}^n$, we will denote by $\mathrm{MAT}(\textbf{v})$ the $m \times n$ matrix whose $i$-th column is $\textbf{vec}(v_i)$. The \textit{rank weight} of $\textbf{v}$ is defined as
$$
\mathrm{rw}(\textbf{v})=\mathrm{rank}(\mathrm{MAT}(\textbf{v}))
$$
and does not depend on the choice of the irreducible polynomial of degree $m$ over $\mathbb{F}_q$. We will denote by $\mathcal{S}(\textbf{v})$ the $\mathbb{F}_q$-vector space spanned by the coordinates of $\textbf{v}$, i.e.,
$$
\mathcal{S}(\textbf{v})= \langle v_1, \ldots, v_n \rangle_{\mathbb{F}_q}.
$$
It is easy to see that $\dim(\mathcal{S}(\textbf{v}))=\mathrm{rw}(\textbf{v})$. Moreover, we will consider
$$
\mathcal{S}^n_w(\mathbb{F}_{q^m})=\{ \textbf{e} \in \mathbb{F}_{q^m}^n : \mathrm{rw}(\textbf{e})=w\}.
$$

Let us consider the map
$$
\mathrm{poly}: (v_0,\ldots,v_{m-1}) \in \mathbb{F}_{q}^m \mapsto \sum_{i=0}^{m-1}v_i X^i \in \mathbb{F}_q[X]/\langle f \rangle
$$
and let us define the product $\cdot$ between two vectors $\textbf{u}, \textbf{v}$ of $\mathbb{F}_{q^m}^n$ in the following way
$$
\textbf{u} \cdot \textbf{v}=\mathrm{poly}^{-1}\Big(\mathrm{poly}(\textbf{u})\mathrm{poly}(\textbf{v})\Big),
$$
where the polynomial product is computed modulo $f$.

Given $\textbf{v} \in \mathbb{F}_q^n$, we denote by $IM_f(\textbf{v})$ the \textit{ideal matrix generated by $\textbf{v}$ modulo $f$} and it is defined as
$$
\begin{pmatrix}  
\textbf{v} \\
\textrm{poly}^{-1}(X\textrm{poly}(\textbf{v}))\\
 \vdots\\
\textrm{poly}^{-1} (X^{n-1}\textrm{poly}(\textbf{v}))
\end{pmatrix}   \in \mathbb{F}_{q}^{n \times n}.
$$
Let $s \geq 2$. An \textit{$s$-ideal code} is an $[sn,n]$-code $\mathcal{C}$, where $n$ is the dimension of $\mathcal{C}$ over $\mathbb{F}$ and $sn$ is the lenght of its vectors, such that its generator matrix has the shape
$$
\textbf{G}=\Big( \textbf{I}_n \,\, IM_f(\textbf{g}_1) \,\, \ldots \,\, IM_f(\textbf{g}_{s-1})  \Big)\in \mathbb{F}_q^{n \times sn},
$$
for some $\textbf{g}_i \in \mathbb{F}_q^n$, $1 \leq i \leq s-1$. Equivalently, an ideal code can be defined by its parity check matrix
$$
\textbf{H}= \begin{pmatrix} 
 & IM_f(\textbf{h}_1)^t  \\
 \textbf{I}_{n(s-1)} & \vdots  \\
  & IM_f(\textbf{h}_{s-1})^t 
\end{pmatrix} \in \mathbb{F}_q^{n(s-1) \times sn},
$$
for some $\textbf{h}_i \in \mathbb{F}_q^n$, $1 \leq i \leq s-1$. In the following, an $s$-ideal code will be equipped with the rank weight.

\section{Homomorphic schemes and coding theory}\label{coding}

In this section, we will describe in detail the homomorphic schemes known in the literature, whose construction is based on coding theory, and we will analyze their security.

\subsection{Armknecht et al.'s scheme}

The following homomorphic scheme was introduced by Armknecht et al. in ~\cite{armknekt}.

Let $\mathcal{X}$ be a geometric object (e.g., $\mathbb{F}$ or $\mathbb{F}^t$) and $\textbf{x}=(x_1,\ldots,x_n)$ be an $n$-tuple of $n$ distinct elements of $\mathcal{X}$. Let $\mathbb{F}^{\mathcal{X}}$ be the set of all functions from $\mathcal{X}$ to $\mathbb{F}$ and let $\mathcal{L} \subset \mathbb{F}^{\mathcal{X}} $ be an $\mathbb{F}$-vector space of dimension $k$. An \textit{evaluation code} $\mathcal{C}$ is obtained through the function $Encode= ev \circ ex$, where
\begin{center}
$ex:\mathbb{F} \longrightarrow \mathcal{L}$, $m \mapsto p$
\end{center}
\begin{center}
$ev:\mathcal{L} \longrightarrow \mathbb{F}^n $, $p \mapsto p(\textbf{x})=(p(x_1),\ldots,p(x_n))$.
\end{center}
We will call $\mathcal{L}(\mathcal{C})=\mathcal{L}$ the \textit{function space} and the vector $\textbf{x}$ the \textit{codeword support}.
Now, let $\mathcal{C}$ and $\tilde{\mathcal{C}}$ be two evaluation codes with the same codeword support such that
\begin{center}
$p_1 \cdot p_2 \in \mathcal{L}(\tilde{\mathcal{C}})$ for every $p_1,p_2 \in \mathcal{L}(\mathcal{C}),$
\end{center}
where $(p_1 \cdot p_2)(x)= p_1(x) \cdot p_2(x)$ for every $x \in \mathcal{X}$.
Then, for every $\textbf{w}_1, \textbf{w}_2 \in \mathcal{C}$ we have $\textbf{w}_1 \cdot \textbf{w}_2 \in \tilde{\mathcal{C}}$. This property can be generalized by introducing the concept of a $\mu$-multiplicative code.

An evaluation code $\mathcal{C}$ is called \textit{$\mu$-multiplicative} if there exists an evaluation code $\tilde{\mathcal{C}}$ with the same codeword support such that $ \mathcal{L}(\mathcal{C})^{\ell} \subset \mathcal{L}(\tilde{\mathcal{C}})$ for every $\ell \in [\mu]$. In the following, we will use the notation $\mathcal{C}^{\mu} \subset \tilde{\mathcal{C}}$.

Therefore, if we use $\mu$-multiplicative codes, we have that the sum and the product of codewords is a codeword. However, this does not imply that the decoding of the product of codewords corresponds to the product of the messages. For this property to be verified, we must introduce the concept of a \textit{special evaluation code}.

An evaluation code $\mathcal{C}$ with codeword support $\textbf{x}=(x_1,\ldots,x_n)$ is called a \textit{special evaluation code} if the function $ex$ is defined as follows. Let $y \in \mathcal{X}$ be such that $y \ne x_i$ for every $i \in [n]$. The element $y$ is called the \textit{message support}. For every $m \in \mathbb{F}$, let
\[\mathcal{L}_{y \mapsto m}=\{ p \in \mathcal{L} : p(y)=m \},\]
and assume that $\mathcal{L}_{y \mapsto m} \ne \emptyset$. Then for every $m \in \mathbb{F}$
\[ex(m)=p \in \mathcal{L}_{y \mapsto m}.\]

A special evaluation code $\mathcal{C}$ \textit{admits a special encoding} of $0 \in \mathbb{F}$ if there exists $\textbf{w}=(w_1,\ldots,w_n) \in \mathcal{C}$ such that $w_i \ne 0$ for every $i=1,\ldots,n$ and $Decrypt(\textbf{w})=0.$

In the following, we will describe the construction of the cryptographic scheme, analyze the security of the proposed scheme, and show a possible implementation based on Reed-Muller codes.

The scheme allows an unlimited number of additions and a fixed but arbitrary number of multiplications. The message space is a field $\mathbb{F}$. If $I \subset [n]$, then
\[
\mathcal{C}(I)=\{ \textbf{w}+\textbf{e} : \textbf{w} \in \mathcal{C}, \textbf{e} \in \mathbb{F}^n \setminus \{ \textbf{0}\}, supp(\textbf{e}) \subset [n] \setminus I \}.
\]
\begin{itemize}
\item \textit{Keygen.} The input parameters are three integers: $s$ the security parameter, $\mu$ the maximum number of multiplications, $L$ the total number of encryptions. Furthermore, the codeword support $\textbf{x}$, the message support $y$, a subset $I \subset [n]$ of cardinality $T$ that represents the good locations, and two special evaluation codes $\mathcal{C}^{\mu} \subset \tilde{\mathcal{C}}$ are determined such that if $n$ denotes the length of the codewords then
\[
n - T \geq L+1 \quad \mathrm{and} \quad n \in O(\mu ^3 s).
\]
The output is the secret key $\textbf{sk}=(\textbf{x},y,I)$.

\item \textit{Encrypt.} The algorithm takes as input the message $m \in \mathbb{F}$ and the secret key $\textbf{sk}=(\textbf{x},y,I)$. Then, it chooses (randomly) a codeword $\textbf{w}$ of the code $\mathcal{C}$, and then (uniformly at random) a vector $\textbf{e} \in \mathbb{F}^n$ such that $supp(\textbf{e}) \subset [n] \setminus I$. The output is given by the pair $(\textbf{c}, \gamma)$, where $\textbf{c}= \textbf{w}+\textbf{e}$ and $\gamma$ counts the number of multiplications performed on $\textbf{c}$.

\item \textit{Decrypt.} The algorithm takes as input the ciphertext $(\textbf{c}, \gamma)$ and the secret key $\textbf{sk}=(\textbf{x},y,I)$. If $\textbf{c}=(c_1,\ldots,c_n) \in \mathcal{C}(I)$ and $\gamma \leq \mu$, then the output is given by
\begin{center}
$p_{\textbf{c}}(y)=m$,
\end{center}
where $p_{\textbf{c}} \in \mathcal{L}(\mathcal{C})$ such that $p_{\textbf{c}}(x_i)=c_i$ for every $i \in I$.

\item \textit{Eval.} Let $(\textbf{c}, \gamma), (\textbf{c}', \gamma')$ be two ciphertexts corresponding to the messages $m_1, m_2,$ respectively.

\textit{Add} $\rightarrow$ $(\textbf{c}+\textbf{c}', \max\{\gamma, \gamma'\})$ and we have $Encrypt(m_1+m_2)=\textbf{c}+\textbf{c}'$.

\textit{Mult} $\rightarrow$ $(\textbf{c} \odot \textbf{c}', \gamma + \gamma')$ with $\gamma + \gamma' \leq \mu,$ and we have $Encrypt(m_1 \cdot m_2)=\textbf{c} \odot \textbf{c}'$.
\end{itemize}
We observe that the scheme is symmetric-key and somewhat homomorphic.

The semantic security of the scheme is based on the difficulty for an attacker, who knows the code, to identify good and bad locations. This problem is called the \textit{Decoding Synchronized Codes Problem (DSCP)}, see \cite[Definition 5]{armknekt}. In particular, the security of the scheme with respect to DSCP is related to the existence of a third code $\overline{\mathcal{C}}$ that admits a special encoding of $0$, and to the choice of specific values for the parameters $n,\rho,q,t,T$ to avoid \textit{Information Set Decoding (ISD)} and \textit{Low Weight Dual Codewords (LWC)} attacks (for more details see \cite{isd} and \cite[Section B.2]{armknekt}, respectively).

Now, let $\mathbb{F}_q$ be a finite field with $q$ elements, $q$ a power of a prime number. Let $\mathcal{X}=\mathbb{F}_q^t,$ $\textbf{x}=(x_1,\ldots,x_n)$ be the codeword support and
\[
\mathcal{L}_{t,\rho}=\{ f \in \mathbb{F}_q[v_1,\ldots,v_t] : \deg(f) < \rho \}.
\]
We denote with
\[RM_q(t,\rho)=\{ (f(x_1),\ldots,f(x_n)) \in \mathbb{F}_q^n : f \in \mathcal{L}_{t, \rho} \}\]
the \textit{Reed-Muller code of order $\rho < q$}. If $n=q^t$, then $RM_q(t, \rho)$ is called \textit{full}. In the implementation of the scheme, we take $n < q^t$ and in this case the code $RM_q(t, \rho)$ is called \textit{punctured}.

Assume that the input parameters $s, \mu, L$ have been assigned. The codeword support
\[
\textbf{x}=((x_1^{(1)},\ldots,x_t^{(1)}), \ldots, (x_1^{(n)},\ldots,x_t^{(n)})) \in (\mathbb{F}_q^t)^n
\]
is determined (randomly) and the message support $y=(y_1,\ldots,y_t) \in \mathbb{F}_q^t$ is chosen such that $y_1 \ne x_1^{(i)}$ for every $i=1,\ldots,n$.

Then, we consider the following Reed-Muller codes
\begin{center}
$\overline{\mathcal{C}}=RM_q(t, \rho),$ $\mathcal{C}=RM_q(t,2 \rho),$ $\tilde{\mathcal{C}}=RM_q(t, 2 \mu \rho)$
\end{center}
having dimensions $\overline{k}=\binom{t + \rho}{t},
k=\binom{t + 2 \rho}{t}, \tilde{k}=\binom{t + 2 \mu \rho}{t}$, respectively. These codes satisfy the following properties:
\begin{itemize}
\item $\mathcal{C}^{\mu} \subset \tilde{\mathcal{C}}$;
\item $\overline{\mathcal{C}}^{2} \subset \mathcal{C}$;
\item $\overline{\mathcal{C}}$ admits a special encoding of $0$.
\end{itemize}
To counter ISD and LWC attacks, we choose the parameters $n, \rho, q, t, T$ as follows:
\[n_{min}= \min_{\rho} \left\{ 2^{\frac{s}{\binom{3+\rho}{\rho}}} \cdot \binom{3+2 \mu \rho}{3} \right\} \leq 2^{\frac{6s}{(\sqrt[3]{6s})^3}} \cdot (3+2 \mu \sqrt[3]{6s})^3,\]
hence $n_{min} \in O(\mu^3s)$,

\[
\rho_{min} \quad \textrm{is the value of $\rho$ for which $n_{min}$ is obtained},
\]

\medskip

\[
q_{min}=\min \left\{ q : \Bigg( \prod_{j=0}^{\rho_{min}+1} \frac{\binom{3+2 \mu \rho_{min}}{3}-j}{q^3-j} \Bigg) \cdot q^2 \cdot (q^2+q+1) \cdot \binom{q}{\rho_{min}+2} \leq 2^{-s} \right\},
\]

\medskip

\[t=3 \quad \mathrm{and} \quad T < q/2.\]

\subsection{Challa - Gunta's schemes}

The following two schemes has been introduced by Challa and Gunta in \cite{challagunta} and \cite{challaguntamatrix}. These schemes are fully homomorphic, symmetric-key and based on Reed-Muller codes. In the following, we denote by $\mathbb{F}_2=\{0,1\}$ a finite field with two elements.

We begin by analyzing the scheme introduced in \cite{challagunta}.

Let $0\leq r \leq m$ be two integers and $n=2^m$. A Reed-Muller code $RM(r,m)$ is defined as the set
\[
RM(r,m)=\{ (f(a_1),\ldots,f(a_{n})) \in \mathbb{F}_2^n : f \in \mathbb{F}_2[x_1,\ldots,x_m], \deg(f)\leq r \},
\]
where $a_1,\ldots,a_{n}$ are all the elements of $\mathbb{F}_2^m$. The dimension $k$ of $RM(r,m)$ and the minimum Hamming distance $d$ are given by
\[
k=\sum_{i=0}^r \binom{m}{i}, \quad d=2^{m-r}.
\]
Let $G_{rm} \in \mathbb{F}_2^{k \times n}$ be a generator matrix of a Reed-Muller code. We have
\[
RM(r,m)=\{ \textbf{v}G_{rm}: \textbf{v} \in \mathbb{F}_2^{k} \}.
\]
\begin{itemize}
\item \textit{Keygen.} Given a security parameter $p$, choose $r,m$ such that $k\geq 2p$ and $n \geq 2k$, where $k$ is the dimension of $RM(r,m)$ and $n$ is the length of the codewords. Let $G_{rm} \in \mathbb{F}_2^{k \times n}$ be a generator matrix whose row vectors are $\textbf{v}_1, \ldots, \textbf{v}_k \in \mathbb{F}_2^n$.

Choose an integer $\ell \geq n^2$.

Choose (randomly) a subset $K \subset [\ell]$ of cardinality $n$.

The output is the secret key $\textbf{sk}=(G_{rm}, K, \ell)$.

\item \textit{Encrypt.} Let $\textbf{m}=(a_1,a_2,\ldots)$ be a message of length $\leq p$.

If the length of $\textbf{m}$ is $t<p$, add $p-t$ zero bits to the left of $\textbf{m}$. In this way, we obtain the vector $\textbf{0} || \textbf{m}$ of length $p$.

Choose (randomly) a vector $\textbf{r}$ of length $(k-p)$.

Add the vector $\textbf{r}$ to the left of $\textbf{m}$ (or $\textbf{0} || \textbf{m}$) to obtain the vector $\textbf{p}=(p_1,p_2,\ldots,p_k)$.

Let $\textbf{w}=p_1\textbf{v}_1+\ldots + p_k \textbf{v}_k=(x_1,\ldots,x_n)$.

Choose (randomly) a vector $\textbf{u}$ of length $\ell$.

The ciphertext $\textbf{c}$ is obtained by replacing the bits of the vector $\textbf{u}$ at the positions specified by the set $K$ with the bits of the vector $\textbf{w}$.

\item \textit{Decrypt.} From the set $K$, the algorithm extracts the vector $\textbf{w}=(x_1,\ldots,x_n)$ from the ciphertext $\textbf{c}$.

Calculate the vector $\textbf{p}=(p_1,\ldots,p_k)$ as follows:
\begin{itemize}
\item[i)] $p_1=x_1$;
\item[ii)] for every $i=2, \ldots, k$, let $p_i=x_1 + x_{2^{i-1}}$.
\end{itemize}

Eliminate the $(k-p)$ bits to the left (most significant bits) of $\textbf{p}$ to obtain the vector $\textbf{0}||\textbf{m}$.

Eliminate the leading zero bits if present (most significant zero bits).

The output is the message $\textbf{m}=(a_1,a_2,\ldots)$.

The identities $p_1=x_1$ and $p_i=x_1 + x_{2^{i-1}}$ hold only when the code is viewed in the standard evaluation form of $RM(1,m)$, where $x_1$ corresponds to evaluation at the zero vector and $x_1 + x_{2^{i-1}}$ corresponds to evaluation at the $i$-th unit vector. In the general $RM(r,m)$ case, these identities recover only the constant and linear coefficients, and a full Reed–Muller decoding procedure is required to recover the remaining higher-degree coefficients.

\item \textit{Eval.} Let $\textbf{c}_1=(x_1,\ldots,x_{\ell})$ and $\textbf{c}_2=(y_1,\ldots,y_{\ell})$ be the ciphertexts corresponding to the messages $\textbf{m}_1$ and $\textbf{m}_2$, respectively.

\textit{Add} $\rightarrow$ $\textbf{c}_3=\textbf{c}_1+\textbf{c}_2$ and we have $\textbf{c}_3=Encrypt(\textbf{m}_1+\textbf{m}_2)$.

\textit{Mult} $\rightarrow$ Let $\textbf{c}_3=(z_i)_{i=1,\ldots,\ell}$ such that
\[
z_i=(x_i \cdot y_i)+(x_i \cdot y_1)+(y_i \cdot x_1).
\]
Then, we have $Encrypt(\textbf{c}_3)=\textbf{m}_1 \odot \textbf{m}_2$.

\end{itemize}

The authors analyze the security of the proposed scheme against the \textit{Indistinguishability Chosen Plaintext (IND--CPA)} attack and other known attacks. Since the encryption algorithm does not add an error term, the confidentiality of the scheme relies on the difficulty of identifying the secret set $K$ that hide the codeword inside the ambient random vector.
\begin{itemize}
\item The scheme is secure against (IND--CPA) and \textit{Brute-force} attacks.
\item \textit{Privacy of the homomorphic operations.} Privacy refers to the fact that the ciphertexts have a fixed length. Therefore, it is difficult to determine the type and number of operations performed on the ciphertexts.
\item If $\ell > n^{5/2}$, the scheme is able to counter attacks against the \textit{DSCP}.
\item Since the function that associates a vector $\textbf{w}$ to each message $m$ is random, it is not possible to identify the set $K$.
\end{itemize}

The next scheme is described in \cite{challaguntamatrix}.

Let $G_{rm} \in \mathbb{F}_2^{k \times n}$ be a generator matrix of a Reed-Muller code with parameters $(r,m)$ and let $\textbf{v}_1,\textbf{v}_2,\ldots,\textbf{v}_k$ be the row vectors. Let $\textbf{m}=(m_{1},m_{2},\ldots,m_{k}) \in \mathbb{F}_2^k$ be a message and let $\textbf{m} \times G_{rm}$ be the $(k \times n)$ matrix whose rows are the vectors $m_1 \textbf{v}_1, \ldots, m_k \textbf{v}_k$. We note that $m_1 \textbf{v}_1+\ldots+m_k \textbf{v}_k=\textbf{m}G_{rm}$ is a codeword of the Reed-Muller code with parameters $(r,m)$.

Consider the code $\mathcal{C}=\{ \textbf{m} \times G_{rm} : \textbf{m} \in \mathbb{F}_2^k \}$. Each matrix $E \in \mathbb{F}_2^{k \times n}$ can be viewed as a $k$-tuple formed by its row vectors, i.e., $E=(\textbf{e}_1,\ldots,\textbf{e}_k)$. Let $I \subset [n]$, then we can consider the set
\[
\mathcal{C}(I)=\{ W+E : W \in \mathcal{C}, E=(\textbf{e}_1,\ldots,\textbf{e}_k) \in \mathbb{F}_2^{k \times n} \setminus \{\textbf{0}\}, supp(\textbf{e}_i) \subset [n] \setminus I \}.
\]

Let $S$ be a permutation on the set $[k] \times [n]$. Let $\sigma_S: \mathbb{F}_2^{k \times n} \longrightarrow \mathbb{F}_2^{k \times n}$ be a function defined as follows: if $W_2=\sigma_S(W_1)$ then
\[
W_2(i,j)=W_1(i',j') \quad \mathrm{with} \quad (i,j)=S(i',j').
\]
We will say that the function $\sigma_S$ is \textit{associated with the permutation key $S$}.

\begin{itemize}

\item \textit{Keygen.} The algorithm chooses the parameters $r,m$ for a Reed-Muller code such that $m \geq 2$ and $0 < r \leq m$.

Choose $S_1 \subset [n]$ a set consisting of bad locations such that $ |S_1| < 2^{m-r-1}$.

Choose a permutation key $S_2$ on the set $[k] \times [n]$.

The output is the secret key $\textbf{sk}=(S_1,S_2)$.

\item \textit{Encrypt.} The algorithm takes as input a message $\textbf{m} \in \mathbb{F}_2^k$ and the secret key $\textbf{sk}=(S_1,S_2)$.

The message $\textbf{m}$ is transformed into the matrix $W=\textbf{m} \times G_{rm} \in \mathcal{C}$.

The algorithm chooses (randomly) a matrix $E=(\textbf{e}_1,\ldots,\textbf{e}_k) \in (\mathbb{F}_2^n)^k \setminus \{\textbf{0}\}$ such that $supp(\textbf{e}_i) \subset S_1$.
The ciphertext $C$ is given by the matrix
\[
C=\sigma_{S_2}(\textbf{m}\times G_{rm} + E).
\]

\item \textit{Decrypt.} The algorithm receives the ciphertext $C$ and the secret key $\textbf{sk}=(S_1,S_2)$. Let $\sigma'$ be the inverse function of $\sigma_{S_2}$, we have
\[
\sigma'(C)=\textbf{m}\times G_{rm}+E=(\textbf{w}_1,\ldots,\textbf{w}_k) \in (\mathbb{F}_2^n)^k.
\]
Let $\textbf{w}=\textbf{w}_1+\ldots+\textbf{w}_k \in \mathbb{F}_2^n$. Since $ |S_1| < 2^{m-r-1}$, by applying the Reed algorithm \cite{reed} to the vector $\textbf{w}$, we obtain the vector $\textbf{m}G_{rm}$, and thus $\textbf{m}$.

\item \textit{Eval.} Let $C_1,C_2 \in \mathbb{F}_2^{k \times n}$ be the ciphertexts corresponding to the messages $\textbf{m}_1$ and $\textbf{m}_2$, respectively.

\textit{Add} $\rightarrow$ $C_3=C_1+C_2$ and we have $C_3=Encrypt(\textbf{m}_1+\textbf{m}_2)$.

\textit{Mult} $\rightarrow$ Let $C_3=(C_3(i,j))$ with
\[
C_3(i,j)=C_1(i,j) \cdot C_2(i,j),
\]
that is $C_3= C_1 \odot C_2$. We have $C_3=Encrypt(\textbf{m}_1 \odot \textbf{m}_2)$.

\end{itemize}

The security of the proposed scheme is based on the set of error positions and the permutation key. During encryption, errors are inserted at positions specified by the set $S_1$, and the resulting erroneous codeword matrix is further masked by a secret permutation $S_2$. Therefore, decryption is straightforward for a legitimate user who knows both the error locations and the permutation, whereas an adversary without this information is expected to face a decoding problem that is intended to resemble decoding a random code.

\subsection{Bogdanov - Lee's scheme}
The following scheme has been introduced by Bogdanov and Lee in \cite{bogdanov} and is somewhat homomorphic and asymmetric-key. We highlight that this scheme is not secure. Indeed, in \cite{attack} the authors show an attack which can decrypt any ciphertext.

The message space is a finite field $\mathbb{F}_q$. 

\begin{itemize}
    \item \textit{Keygen.} Let $n$ be the security parameter. The algorithm chooses (uniformly at random) a subset $S \subset [n]$ of cardinality $s$ and a matrix $M$ of order $n \times r$ as follows. First, $n$ distinct elements $a_1, \ldots, a_n \in \mathbb{F}_q$ are chosen (uniformly at random). The $i$-th row $M_i$ is defined as
    \[
    M_i= \begin{cases} (a_i, a_i^2, \ldots, a_i^{s/3},0, \ldots,0) & \mbox{if } i\in S,\\ (a_i, a_i^2, \ldots, a_i^{s/3},a_i^{s/3+1},\ldots,a_i^r) & \mbox{if } i\not\in S. \end{cases}
    \]
    The secret key is given by the pair $\textbf{sk}=(S,M)$ and the public key by the matrix $\textbf{pk}=P=MR$, where $R$ is a matrix of order $r \times r$ over $\mathbb{F}_q$ with determinant equal to one.

    The matrices $P,M$ are generator matrices for the same linear code.

    \item \textit{Encrypt.} Let $m \in \mathbb{F}_q$ be a message and $P$ be the public key. The algorithm chooses (uniformly at random) a vector $\textbf{x} \in \mathbb{F}_q^r$ and an error vector $\textbf{e}=(e_1,\ldots,e_n) \in \mathbb{F}_q^n$ by choosing each $e_i$ from a distribution $\mathcal{X}$ over $\mathbb{F}_q$.

    The output is the vector $\textbf{c}=P\textbf{x}+m\textbf{1}+\textbf{e} \in \mathbb{F}_q^n$, where $\textbf{1}=(1,\ldots,1) \in \mathbb{F}_q^n$.

    \item \textit{Decrypt.} Given the secret key $\textbf{sk}=(S,M)$, to decrypt the ciphertext $\textbf{c}$, the algorithm finds a solution to the following system of $s/3+1$ linear equations in the variables $y_i \in \mathbb{F}_q$, $i \in S$:
    \begin{equation}\label{sistemabogdanov}
    \begin{cases} \sum_{i \in S} y_i M_i= \textbf{0} \\ \sum_{i \in S}y_i=1. \end{cases}
    \end{equation}
    The output is given by $\sum_{i \in S}y_ic_i$.

    The $Decrypt$ algorithm returns the message $m$ as output if $e_i = 0$ for every $i \in S$.
    \item \textit{Eval.} Let $\textbf{c}_1$, $\textbf{c}_2$ be the ciphertexts corresponding to the messages $\textbf{m}_1$ and $\textbf{m}_2$, respectively.

    \textit{Add} $\rightarrow$ $\textbf{c}_3=\textbf{c}_1+\textbf{c}_2$ and we have $\textbf{c}_3=Encrypt(\textbf{m}_1+\textbf{m}_2)$.

    \textit{Mult} $\rightarrow$ $\textbf{c}_3=\textbf{c}_1 \odot \textbf{c}_2$ and we have $\textbf{c}_3=Encrypt(\textbf{m}_1 \cdot \textbf{m}_2)$.
\end{itemize}

Let $\eta = \mathbb{P}[\mathcal{X} \ne 0]$, $\alpha \in (0, \frac{1}{4}]$ and $M_S$ be the submatrix of $M$ whose rows are indexed by the set $S$. The authors analyze the security of the scheme from the perspective of the following attacks:
\begin{itemize}
    \item \textit{Recover the subset $S$ from the public key $P$.} The attack requires exponential time if $s = n^{\alpha/4}$ and $q \geq 2^{n^{\alpha}}$.
    \item An adversary can consider a system similar to \eqref{sistemabogdanov}:
    \begin{equation}\label{sistemaattacco}
    \begin{cases} \sum_{i \in [n]} y_i P_i= \textbf{0} \\ \sum_{i \in [n]}y_i=1. \end{cases}
    \end{equation}
    The set of solutions of \eqref{sistemaattacco} does not change if we consider the matrix $M$, so \eqref{sistemaattacco} contains the solutions of \eqref{sistemabogdanov} (with $y_i=0$ if $i \not \in S$). If the system \eqref{sistemaattacco} admits a unique solution, the adversary will be able to decrypt the messages. By taking $r$ smaller than $n$, for example setting $r = n^{1-\alpha/8}$, the system \eqref{sistemaattacco} has more than one solution and many of these have large Hamming distance. These solutions are not useful for decrypting messages.
    \item \textit{Recover the vector $\textbf{x}$ used in the encryption algorithm}. The attack requires exponential time if $\eta=1/n^{1-\alpha/4}$.
\end{itemize}

\textbf{Conjecture.} For every $\alpha \in (0, \frac{1}{4}]$ there exists $\gamma > 0$ such that the cryptographic scheme with parameters
$r = n^{1-\alpha/8}, \eta = 1/n^{1-\alpha/4}, s = n^{\alpha/4}, q \geq 2^{n^{\alpha}}$ is \textit{$(2^{n^{\gamma}},2^{-n^{\gamma}})$-message indistinguishable}, for every $n$
sufficiently large.\\

The Bogdanov--Lee cryptosystem was broken by Gauthier, Otmani, and Tillich through a distinguisher-based attack relying on the algebraic structure of the public code. The key observation is that the public code is not a random linear code, but a modified Reed--Solomon code obtained by inserting a zero submatrix into the Vandermonde matrix defining the code. The columns corresponding to this modification form a secret set, denoted by $S$. Once this set is recovered, any ciphertext can be decrypted.

The attack is based on the \emph{square code}. If $\mathcal{C}$ is a linear code, its square code $\langle \mathcal{C}^2 \rangle$ is the linear span of all component-wise products of pairs of codewords of $\mathcal{C}$. Reed--Solomon codes have an unusually small square code dimension compared with random linear codes, and this abnormal behaviour can be used as a distinguisher. In the Bogdanov--Lee setting, one considers punctured versions of the public code: for a subset $I$ of coordinates, let $\mathcal{C}_I$ be the restriction of the public code to the positions in $I$. The crucial fact proved in~\cite{attack} is that the dimension of the punctured square code $\langle \mathcal{C}_I^2 \rangle$ depends directly on the number of secret positions contained in $I$.

More precisely, if $I$ is a subset of coordinates and $J=I\cap S$, then, under suitable size conditions on $I$ and $J$, the paper shows that
\[
\dim\big(\langle \mathcal{C}_I^2 \rangle\big)=2k-1+|J|,
\]
where $k$ is the dimension of the public code. Therefore, the dimension of the square code increases exactly with the number of secret coordinates of $S$ contained in $I$. This yields an efficient distinguisher: by comparing the dimensions of $\langle \mathcal{C}_I^2 \rangle$ and $\langle \mathcal{C}_{I'}^2 \rangle$ for sets $I'$ obtained by removing or replacing coordinates of $I$, one can test whether a given position belongs to $S$.

The attack proceeds in two phases. First, one chooses a subset $I$ of coordinates and computes the dimension of $\langle \mathcal{C}_I^2 \rangle$. Then one removes a coordinate $x\in I$ and recomputes the dimension. If the dimension decreases by one, then $x\in S$; otherwise, $x\notin S$. Repeating this test allows the adversary to recover all positions of $I\cap S$. In a second phase, the attacker exchanges known non-secret positions in $I$ with coordinates outside $I$, and again observes the variation in the square code dimension. In this way, the whole secret set $S$ can be recovered in polynomial time.

Once $S$ is known, the scheme can be completely broken. Indeed, the attacker can solve the same linear system used in the secret decryption procedure, but with the public matrix $P$ in place of the secret matrix $M$. Since $P=MR$ with $R$ invertible, the condition $Py^t=0$ is equivalent to $My^t=0$. Hence the adversary can reconstruct a valid vector $y$ supported on $S$ and decrypt any ciphertext exactly as the legitimate receiver does. The attack is therefore a full key-recovery attack and shows that the Bogdanov--Lee scheme is insecure.

\subsection{Aguilar-Melchor et al.'s scheme}

The following scheme has been introduced by Aguilar-Melchor et al. in \cite{random}. This scheme is symmetric-key and somewhat homomorphic  based on random rank metric ideal linear codes. 

Let $\mathbb{F}_q$ be a finite field with $q$ elements. Let $m,n, w$ be the degree extension of $\mathbb{F}_q$, the length of the vectors and  the rank weight of the errors such that $\frac{w(w+3)}{2}+1<m$, respectively.

\begin{itemize}
    \item \textit{Keygen}. The algorithm chooses uniformly $g_1 \in \mathbb{F}_{q^m}$, $\textbf{x}=(x_1,\ldots,x_w) \in \mathcal{S}_w^w(\mathbb{F}_{q^m})$. Let $\textbf{X}=\mathcal{S}(\textbf{x})$ and $\bar{\textbf{X}}=\mathcal{S}(\textbf{x}, g_1\textbf{x}, (f_if_j)_{1 \leq i,j\leq w})$. Then, it computes a basis $\textbf{a}=(a_1,\ldots,a_{d}) \in \mathcal{S}_d^d(\mathbb{F}_{q^m})$ of $\bar{\textbf{X}}$, where $d=\dim \bar{\textbf{X}}$. Now, it puts $g_2=g_1^2$ and checks if $\mathrm{rw}(a_1,\ldots,a_d,g_1,g_2)=d+2$ (if not, then the algorithm restarts from the beginning). Thus, the vector $(a_1,\ldots,a_d,g_1,g_2)$ is extended to a basis $\textbf{b}=(a_1,\ldots,a_d,g_1,\ldots,g_{m-d})$ of $\mathcal{S}_m^m(\mathbb{F}_{q^m})$. Put $\textbf{g}=(g_1,\ldots, g_{m-d})$ and compute the matrix $\textbf{B}=\mathrm{Mat}(\textbf{b})$. Define the matrix $\textbf{D}$ made up by the last  $m-d$ columns of $(\textbf{B}^{-1})^t$. Finally, it chooses uniformly $\textbf{s} \in \textbf{X}^n$ and the secret key is given by $\textbf{sk}=(\textbf{x},\textbf{g}, \textbf{D}, \textbf{s})$.

    \item \textit{Encrypt}. Let $\textbf{m} \in \mathbb{F}_q^n$ be a message. Let $\textbf{r}=(\textbf{r}_1, \textbf{R}_2) \in \mathbb{F}_{q^m}^n \times \mathbb{F}_q^{w \times n}$ and $\textbf{e}=\textbf{x}\textbf{R}_2$. The algorithm computes $\textbf{v}=\textbf{s} \cdot \textbf{r}_1+\textbf{e}+\tilde{\textbf{m}}$, where $\tilde{\textbf{m}}=g_1 \textbf{m} \in \mathbb{F}_{q^m}^n$, and outputs the chipertext $\textbf{ct}=(\textbf{r}_1,\textbf{v})$.

    \item \textit{Decrypt}. Let $\textbf{d}$ be the first columns of the matrix $\textbf{D}$. The algorithm outputs the message $\textbf{m}=\textbf{d}^t\mathrm{Mat}(\textbf{v}-\textbf{s}\cdot \textbf{r}_1)$.

    \item \textit{Eval}. Let $\textbf{ct}=(\textbf{u}, \textbf{v})$ and $\textbf{ct}'=(\textbf{u}', \textbf{v}')$ be the ciphertexts corresponding to the messages $\textbf{m}$ and $\textbf{m}'$, respectively. 

    \textit{Add} $\rightarrow$ The sum between $\textbf{ct}$ and $\textbf{ct}'$ is defined as $(\textbf{u}+\textbf{u}', \textbf{v}+\textbf{v}')$.

    \textit{Mult} $\rightarrow$ The product between $\textbf{ct}$ and $\textbf{ct}'$ is defined as the triple $(\textbf{a},\textbf{b},\textbf{c}) \in \mathbb{F}_{q^m}^n \times \mathbb{F}_{q^m}^n \times \mathbb{F}_{q^m}^n$, where $\textbf{a}=\textbf{v} \cdot \textbf{v}', \textbf{b}=-(\textbf{u} \cdot \textbf{v}'+\textbf{u}' \cdot \textbf{v}), \textbf{c}=\textbf{u} \cdot \textbf{u}'$. Let $\textbf{d}'$ be the second column of the matrix $\textbf{D}$ and $\textbf{t}=\textbf{a}+\textbf{s} \cdot \textbf{b}+\textbf{s}\cdot \textbf{s}\cdot \textbf{c}$. Then, $\textbf{m} \cdot \textbf{m}'=\textbf{d}'^t \mathrm{Mat}(\textbf{t})$.

    This scheme supports also the multiplication between a message and a ciphertext, called \textit{plaintext absorption}.

    \textit{PtMult} $\rightarrow$ Let $\textbf{m}'' \in \mathbb{F}_{q}^n$. The product between $\textbf{m}''$ and $\textbf{ct}$ is defined as the ciphertext $(\textbf{m}'' \cdot \textbf{u}, \textbf{m}'' \cdot \textbf{v})$.
\end{itemize}

An unlimited number of additions and plaintext absorptions are allowed, but one ciphertext multiplications. The scheme can be also viewed as an additive scheme that supports only addition and plaintext absorption. In this case, the Keygen algorithm has a different version and it is described below, and the bound on the rank weight of the errors is $w<m$. The Encrypt, Decrypt and Evaluation algorithms remain the same.

\begin{itemize}
\item \textit{Keygen} of the additive scheme. The algorithm chooses uniformly $\textbf{x}=(x_1,\ldots,x_w) \in \mathcal{S}_w^w(\mathbb{F}_{q^m})$. Then $\textbf{x}$ is extended to a basis $\textbf{b}=(x_1,\ldots,x_w,g_1,\ldots,g_{m-w})$ of $\mathcal{S}_m^m(\mathbb{F}_{q^m})$. Let $\textbf{g}=(g_1,\ldots,g_{m-w})$ and compute the matrix $\textbf{B}=\mathrm{Mat}(\textbf{b})$.  Define the matrix $\textbf{D}$ made up by the last  $m-w$ columns of $(\textbf{B}^{-1})^t$.  Finally, it chooses uniformly $\textbf{s} \in \textbf{X}^n$, with $\textbf{X}=\mathcal{S}(\textbf{x})$, and the secret key is given by $\textbf{sk}=(\textbf{x},\textbf{g}, \textbf{D}, \textbf{s})$.
\end{itemize}

The security of the scheme is based on the \textit{Ideal Rank Metric Decoding} (IRSD) problem, which can be stated as the following: given any integers $n,w,s$, a random parity check matrix $\textbf{H}$ of an $s$-ideal code $\mathcal{C}$ and $\textbf{y}$ chosen uniformly at random in $\mathbb{F}_{q^m}^{sn-n}$, the problem is to find $\textbf{x} \in \mathbb{F}_{q^m}^{sn}$ such that $\textrm{rw}(\textbf{x})=w$ and $\textbf{y}=\textbf{x}\textbf{H}^t$. The problem of decrypting $\ell$ independent ciphertexts is equivalent to the IRSD problem for an $(l+1)$-ideal code, and for small values of $\ell$ this problem is hard.

The IRSD problem can be solved in polynomial time if the number of published independent ciphertexts is greater than or equal to $2w$, see \cite[Proposition 8 and Lemma 1]{random}. The authors provided also bootstrapping algorithms that makes the scheme fully homomorphic. However, the proposed algorithms are not secure since the number of bootstrapping keys is greater than $2w$. We highlight that this is a limitation of the specific bootstrapping strategy currently proposed and establishing whether secure bootstrapping can be achieved in this setting remains an important open problem.

\section{Homomorphic schemes and polynomials}

In this section, we will describe homomorphic cryptographic schemes whose construction is based on polynomials.

\subsection{Dasgupta - Pal's scheme}
The scheme introduced in \cite{polynomialbit} is symmetric-key and somewhat homomorphic. Through a \textit{refresh} procedure, the scheme becomes fully homomorphic.

Let $\ell$ be the security parameter. The message space is $\mathbb{Z}_N$. Let $m \in \mathbb{Z}_N$ be a message and let $n+1$ be the bit length. Then $m$ will be seen as a polynomial $m(x)$ of degree $n$ whose coefficients are the bits representing $m$.

\begin{itemize}
    \item \textit{Keygen.} The algorithm determines a prime number $S_k$ with $\ell$ bits. The output is the secret key $\textbf{sk}=S_k$.

    It chooses an integer $z$ with $\gamma = \log_2 \ell$ bits.

    Let $R_k = z \cdot S_k$ be the key used in the \textit{Refresh} algorithm.

    \item \textit{Encrypt.} Let $m(x)$ be a message. The algorithm chooses a polynomial $y(x)$ of degree $n$ such that
    \[
    m(x) \equiv y(x) \pmod{S_k}.
    \]

    Given an integer $a$, the algorithm chooses a polynomial $d(x)$ of degree $n$ such that its coefficients are integers with $\ell^a$ bits.

    The output is the ciphertext $c(x)=y(x)+S_k \cdot d(x)$.

    \item \textit{Decrypt.} Given the secret key $S_k$ and the ciphertext $c(x)$, the algorithm has as output
    \[
    (c(x) \pmod{S_k}) \pmod{2} = m(x).
    \]

    \item \textit{Refresh.} Performing operations on ciphertexts can produce an error such that the $Decrypt$ algorithm does not return the correct message. To eliminate the produced error, it is necessary to calculate
    \[c'(x) \equiv c(x) \pmod{R_k},\] and then apply the $Decrypt$ algorithm to the text $c'(x)$.

    \item \textit{Eval.} Let $p_1(x), p_2(x)$ be the ciphertexts corresponding to the messages $m_1, m_2$, respectively.

    \textit{Add} $\rightarrow$ $p_3(x)=p_1(x)+p_2(x)$ and we have $Encrypt(m_1+m_2)=p_3(x).$

    \textit{Mult} $\rightarrow$ $p_3(x)=p_1(x) \cdot p_2(x)$ and we have $Encrypt(m_1 \cdot m_2)=p_3(x).$
\end{itemize}

The security of the scheme is more accurately related to the hardness of the Approximate Greatest Common Divisor (AGCD) problem, rather than to the hardness of factoring large integers as stated in \cite{polynomialbit}. In particular, the ciphertext coefficients can be viewed as noisy multiples of the secret prime $S_k$, and security relies on the difficulty of recovering $S_k$ from such approximate multiples.

\subsection{Dowerah - Krishnaswamy's scheme}
In \cite{multpol}, the authors introduced a symmetric-key and somewhat homomorphic scheme based on polynomials in multiple indeterminates.

Let $I$ be an ideal of the polynomial ring $\mathbb{F}_q[x_1,\ldots,x_{\ell}]$ and $I_{\leq r} = \{f \in I : \deg(f) \leq r \}$. Let
\[
N=\dim_q \mathbb{F}_q[x_1,\ldots,x_{\ell}]_{\leq r}=\binom{{\ell}+r}{{\ell}}.
\]
We assume $q^\ell \geq N$. A polynomial $f \in \mathbb{F}_q[x_1, \ldots , x_{\ell}]_{\leq r}$ evaluated at all points of $\mathbb{F}_q^{\ell}$ determines a vector of $\mathbb{F}_q^{q^{\ell}}$. The set of all vectors obtained by evaluating all polynomials in
$\mathbb{F}_q[x_1, \ldots , x_{\ell}]_{\leq r}$ forms an $N$-dimensional subspace of $\mathbb{F}_q^{q^{\ell}}$. By evaluating all polynomials in $I_{\leq r}$ we obtain a subspace of dimension
$\alpha=\dim_q(I_{\leq r})$ and we will denote it by $\mathcal{V}_{I_{\leq r}}$.

We will denote by $n$ the length of the ciphertexts with $n < q$ and $\alpha<n \leq N$. We choose $n$ distinct points $\textbf{z}_1, \ldots, \textbf{z}_n \in \mathbb{F}_q^{{\ell}}$ such that:
\begin{itemize}
    \item[$i)$] every vector in $\mathbb{F}_q^n$
    can be obtained by evaluating a polynomial in $\mathbb{F}_q[x_1,\ldots , x_{\ell}]_{\leq r}$ at the points $(\textbf{z}_1, \ldots , \textbf{z}_n)$;
    \item[$ii)$] every vector in $\mathbb{F}_q^{\alpha}$
    can be obtained by evaluating a polynomial in $I_{\leq r}$ at the points $(\textbf{z}_1, \ldots , \textbf{z}_{\alpha})$.
\end{itemize}

Let $\textbf{v}_i \in \mathbb{F}_q^N$ be the vector obtained by evaluating the monomials of degree $\leq r$ at the point $\textbf{z}_i$, for every
$1 \leq i \leq n$. Let $G \in \mathbb{F}_q^{n \times N}$ be the matrix whose row vectors are $\textbf{v}_1, \ldots, \textbf{v}_n$.

In the following, we will consider $\mathcal{V}_{I_{\leq r}}$ as a subspace of $\mathbb{F}_q^{n}$ and denote by $(\mathcal{V}_{I_{\leq r}})^{\perp}$ its orthogonal complement. Every vector $\textbf{s} \in (\mathcal{V}_{I_{\leq r}})^{\perp}$ will be represented as
\[
\textbf{s}=(\textbf{s}_1,\textbf{s}_2),
\]
with $\textbf{s}_1$ and $\textbf{s}_2$ having lengths $\alpha$ and $n-\alpha$, respectively.

Let $\mathcal{X}$ be a discrete Gaussian distribution with zero mean over $\mathbb{Z}_q$.

\begin{itemize}
    \item \textit{Keygen.} Let $\lambda$ be the security parameter and $q$ be polynomial in $\lambda$. Let $I \subset \mathbb{F}_q[x_1, \ldots , x_{\ell}]$ be an ideal
    and $\textbf{z}_1, \ldots , \textbf{z}_n \in \mathbb{F}_q^{\ell}$ satisfying conditions $i)$ and $ii)$. Then, let $G \in \mathbb{F}_q^{n\times N}$ be the matrix obtained with the points $\textbf{z}_1, \ldots , \textbf{z}_n$.

    Let $\textbf{s}=(\textbf{s}_1,\textbf{s}_2)=(u_1,\ldots,u_n) \in (\mathcal{V}_{I_{\leq r}})^{\perp}$ and $p \in \mathbb{N}\setminus \{0\}$ such that, if $\sigma_{\textbf{s}}= \sum_{i=1}^n u_i,$ then $\sigma_{\textbf{s}} \cdot p < \lfloor q/2 \rfloor$.

    The secret key consists of a basis of the ideal $I$ and the points $\textbf{z}_1, \ldots, \textbf{z}_n$.

    \item \textit{Encrypt.} Let $m \in \{0, 1\}$ be a message. The algorithm chooses (uniformly at random) a polynomial $f$ of degree $\leq r$ in the ideal $I$ and a vector $\textbf{e} = (\textbf{0}, \overline{\textbf{e}})$, where $\overline{\textbf{e}}$ is a vector of length $n-\alpha$ and the coordinates of the vector $\overline{\textbf{e}}$ are chosen according to the distribution
    $\mathcal{X}$. If $\textbf{f}$ is the vector of coefficients of the polynomial $f$, $\textbf{1}$ has length $n$ and $M = m\cdot p$, then the output is the ciphertext
    \[
    \textbf{c} = M \cdot \textbf{1} + G \cdot \textbf{f} + \textbf{e} \pmod{q}.
    \]
    We note that $G \cdot \textbf{f} \in \mathbb{F}_q^n$ coincides with the vector obtained by evaluating $f$ at the points $\textbf{z}_1, \ldots, \textbf{z}_n$.

    \item \textit{Decrypt.} Let $\textbf{c}$ be a ciphertext. The algorithm returns as output the message
    \[
    m= \Bigg\lfloor \frac{1}{\sigma_{\textbf{s}} \cdot p} \big( \langle \textbf{s} , \textbf{c} \rangle \mod{q} \Big) \Bigg\rfloor \pmod{2},
    \]
    where $\langle \, , \, \rangle$ denotes the standard scalar product.

    \item \textit{Eval.} Let $\textbf{c}_1$ and $\textbf{c}_2$ be the ciphertexts corresponding to the messages $\textbf{m}_1$ and $\textbf{m}_2$, respectively.

    \textit{Add} $\rightarrow$ let $\textbf{c}_{3}=\textbf{c}_1+\textbf{c}_2 \pmod{q}$, the algorithm $Decrypt$ correctly returns the message $m_1+m_2 \pmod{2}$ if
    \begin{center}
    $(m_1+m_2)\cdot \sigma_{\textbf{s}} p < \lfloor q/2 \rfloor$ and $|e_{3}|< \lfloor (\sigma_{\textbf{s}}\cdot p)/2 \rfloor$,
    \end{center}
    with $e_{3}=\langle \textbf{s}_2, \overline{\textbf{e}_1}+\overline{\textbf{e}_2} \rangle \pmod{q}$.

    \textit{Mult} $\rightarrow$ let $\textbf{c}_{3}=\frac{1}{p} (\textbf{c}_1 \odot \textbf{c}_2) \pmod{q}$, the algorithm $Decrypt$ correctly returns the message $m_1 \cdot m_2 \pmod{2}$ if
    \[
    |e_{3}|=|\langle \textbf{s}_2, m_1 \cdot \overline{\textbf{e}_1}+m_2 \cdot \overline{\textbf{e}_2}+ \frac{1}{p} \cdot (\overline{\textbf{e}_1} \odot \overline{\textbf{e}_2}) \rangle| < \lfloor (\sigma_{\textbf{s}}\cdot p)/2 \rfloor.
    \]
\end{itemize}

The scheme is able to counter an \textit{Indistinguishability Chosen Plaintext (IND-CPA)} attack.

\subsection{BFV scheme}
The Brakerski--Fan--Vercauteren (BFV) scheme has been introduced in \cite{bfv} and is an asymmetric-key and leveled fully homomorphic scheme. 

Given a real number $a$, we denote by $\lfloor a \rfloor$ and $\lceil a \rfloor$, respectively, the floor of $a$ and the nearest integer to $a$.

Given an integer $q$ and $n$ a power of 2, let $\mathcal{R}_q=\mathbb{Z}_q[x]/(x^n+1)$. We denote by $\mathcal{X}$ a distribution over $\mathcal{R}=\mathbb{Z}[x]/(x^n+1)$, and let $p<q$ be an integer and $\Delta= \lfloor q/p \rfloor$.

\begin{itemize}
    \item \textit{Keygen.} Let $\lambda$ be the security parameter and $q$ an integer polynomial in $\lambda$. The algorithm outputs a secret key $\textbf{sk}$ and a public key $\textbf{pk}$.

    The secret key $\textbf{sk}=s \in \mathcal{R}$ is chosen according to the distribution $\mathcal{X}$.

    The public key $\textbf{pk}$ is given by the pair
    \[\textbf{pk}=(-(a \cdot s+e) \mod{q}, a),\]
    where $a \in \mathcal{R}_q$ is chosen from a uniform distribution and $e \in \mathcal{R}$ is chosen from the distribution $\mathcal{X}$.

    \item \textit{Encrypt.} Let $m \in \mathcal{R}_p$ be a message and let $\textbf{pk}=(\textbf{pk}[0],\textbf{pk}[1])$ be the public key. The algorithm chooses three elements $u, e_1, e_2 \in \mathcal{R}$ from the distribution $\mathcal{X}$ and outputs the ciphertext
    \[
    \textbf{ct}=(\textbf{pk}[0]\cdot u+ \Delta \cdot m + e_1 \mod{q}, \textbf{pk}[1]\cdot u + e_2 \mod{q}).
    \]

    \item \textit{Decrypt.} Let $\textbf{ct} = (\textbf{ct}[0], \textbf{ct}[1])$ be the ciphertext. The algorithm outputs the message
    \[
    m = \Bigg\lceil \frac{ (\textbf{ct}[0]+ s \cdot \textbf{ct}[1] \mod{q})}{\Delta} \Bigg\rfloor \mod{p}.
    \]

    \item \textit{Eval.} To show the homomorphic properties, a ciphertext $\textbf{ct} = (\textbf{ct}[0], \textbf{ct}[1])$ is viewed as a polynomial of degree 1 in the indeterminate $s$, i.e.,
    \[
    \textbf{ct}(s)=\textbf{ct}[0]+s \cdot \textbf{ct}[1].
    \]
    Let $\textbf{ct}_1(s) = \textbf{ct}_1[0]+ s \cdot \textbf{ct}_1[1]$ and $\textbf{ct}_2(s) = \textbf{ct}_2[0]+ s \cdot \textbf{ct}_2[1]$ be the ciphertexts corresponding to the messages $m_1$ and $m_2$, respectively.

    \textit{Add} $\rightarrow$ $\textbf{ct}_3(s) = \textbf{ct}_1(s) + \textbf{ct}_2(s)$ and we have $Encrypt(m_1+m_2)=\textbf{ct}_3(s).$

    \textit{Mult} $\rightarrow$ We note that
    \[
    \textbf{ct}_1(s) \cdot \textbf{ct}_2(s)= c_0 + c_1 \cdot s + c_2 \cdot s^2,
    \]
    and thus if $c_2 \ne 0$, then $\textbf{ct}_1(s) \cdot \textbf{ct}_2(s)$ is a polynomial of degree $2$. To be able to apply the $Decrypt$ algorithm, it is necessary to reduce its degree to 1. At this point, we introduce a new key $\textbf{rlk}$, called \textit{relinearization key}. Given the secret key $\textbf{sk}=s$ and an integer $\ell$, the key $\textbf{rlk}=(\textbf{rlk}[0], \textbf{rlk}[1])$ is given by the pair
    \[
    \textbf{rlk}=(-(a \cdot s +e)+\ell \cdot s^2 \mod{(q \cdot \ell)}, a),
    \]
    where $a \in \mathcal{R}_{q \cdot \ell}$ is chosen from a uniform distribution and $e \in \mathcal{R}$ is chosen from a distribution $\mathcal{X}' \ne \mathcal{X}$ over $\mathcal{R}$.

    The following values are determined:
    \[
    c_0= \Bigg\lceil \frac{\textbf{ct}_1[0] \cdot \textbf{ct}_2[0]}{\Delta} \Bigg\rfloor \mod{q},
    \]
    \[
    c_1= \Bigg\lceil \frac{\textbf{ct}_1[0] \cdot \textbf{ct}_2[1]+\textbf{ct}_1[1] \cdot \textbf{ct}_2[0]}{\Delta} \Bigg\rfloor \mod{q},
    \]
    \[
    c_2= \Bigg\lceil \frac{\textbf{ct}_1[1] \cdot \textbf{ct}_2[1]}{\Delta} \Bigg\rfloor \mod{q},
    \]
    \[
    (c_{2,0},c_{2,1})=\Bigg( \Bigg\lceil \frac{c_2 \cdot \textbf{rlk}[0]}{\ell} \Bigg\rfloor  \mod{q}, \Bigg\lceil \frac{c_2 \cdot \textbf{rlk}[1]}{\ell} \Bigg\rfloor \mod{q} \Bigg).
    \]
    This algorithm outputs $$\textbf{ct}_3(s)=(c_0+c_{2,0})+s \cdot (c_1+c_{2,1}) \pmod{q}.$$ We have $Encrypt(m_1 \cdot m_2)=\textbf{ct}_3(s)$.
\end{itemize}

The proposed scheme is secure against an \textit{Indistinguishability Chosen Plaintext (IND-CPA)} attack.
To describe the security of the proposed scheme, the authors follow the analysis of Lindner and Peikert in \cite{lindner}. Let $\sigma^2$ be the variance of the distribution $\mathcal{X}$ and let $\delta$ be such that
\[
\log_2(\delta)= 1.8/(\lambda + 100).
\]
Any attack described in \cite{lindner} to succeed with advantage $\varepsilon$,
it is necessary to find vectors of length $\alpha \cdot(q/\sigma)$, with $\alpha=\sqrt{\log_e(1/\varepsilon)/\pi}$. For a fixed $\delta$, Lindner and
Peikert in \cite{lindner} show that, using an optimal attack strategy, the length of the shortest vector that can be computed is given by $2^{2 \sqrt{n \log_2(q)\log_2(\delta)}}$. This implies that the parameters $q, \sigma, n$ must be chosen considering the following relation:
\[
\alpha \cdot \frac{q}{\sigma } < 2^{2 \sqrt{n \log_2(q)\log_2(\delta)}}.
\]

\subsection{BGV scheme}


The Brakerski--Gentry--Vaikuntanathan (BGV) scheme has been introduced in \cite{BGV} and is an asymmetric-key and leveled fully homomorphic scheme.  Let
\[
\mathcal{R} = \mathbb{Z}[x]/(x^n+1),
\]
where $n$ is a power of two, and let
\[
\mathcal{R}_q =\mathbb{Z}_q[x]/(x^n+1), \qquad \mathcal{R}_t = \mathbb{Z}_t[x]/(x^n+1).
\]
 The plaintext space is defined over $\mathcal{R}_t$, whereas ciphertexts are elements of $\mathcal{R}_q^2$ (or, more generally, tuples in $R_q$ whose size may increase after multiplication). Let $\mathcal{X}$ be an error distribution over $\mathcal{R}$.

 \begin{itemize}
     \item \textit{Keygen.} Let $\lambda$ be the security parameter. The algorithm outputs a secret key \textbf{sk} and a public key \textbf{pk}. 
     
     The secret key \textbf{sk} is sampled uniformly at random as an element $s \in \mathcal{R}_q$.
     
     The public key \textbf{pk} is the pair
     $$
     (b,a)=(-a \cdot s + t \cdot e \pmod{q}, a),
     $$
where $e$ is sampled from the distribution $\mathcal{X}$.
     \item \textit{Encrypt.} Let $m \in \mathcal{R}_t$ be a message. The algorithm samples $u,e_1,e_2$ from the distribution $\mathcal{X}$ and computes
\[
c_0 = bu + te_1 + m \pmod q, \qquad
c_1 = au + te_2 \pmod q.
\]
The ciphertext is
\[
\mathbf{c}=(c_0,c_1)\in R_q^2.
\]

     \item \textit{Decrypt.} Given a ciphertext $\mathbf{c}=(c_0,c_1)$, decryption is performed by computing
\[
v = c_0 + c_1 s \pmod q.
\]
Substituting the definition of the public key yields
\[
v = m + t(eu + e_1 + e_2 s) \pmod q.
\]
Hence, reducing $v$ modulo $t$ correctly recovers the plaintext:
\[
m = v \bmod t.
\]
 \item \textit{Eval.} To show the homomorphic properties, a ciphertext $\mathbf{c} = (c_0,c_1)$ is viewed as a polynomial of degree 1 in the indeterminate $s$, i.e.,
    \[
    \mathbf{c}(s)=c_0+s \cdot c_1.
    \]
     Let $\mathbf{c}_1(s) = c_{10}+ s \cdot c_{11}$ and $\mathbf{c}_2(s) = c_{20}+ s \cdot c_{21}$ be the ciphertexts corresponding to the messages $m_1$ and $m_2$, respectively.

    \textit{Add} $\rightarrow$ $\mathbf{c}_3(s) = \mathbf{c}_1(s) + \mathbf{c}_2(s)$ and we have $Encrypt(m_1+m_2)=\mathbf{c}_3(s).$

    \textit{Mult} $\rightarrow$ We note that
    \[
    \mathbf{c}_1(s)\cdot \mathbf{c}_2(s) = d_0 + d_1 \cdot s + d_2 \cdot s^2
    \]
    is a polynomial of degree 2 in $s$. By applying the relinearization algorithm introduced in the BFV scheme, the degree of $\mathbf{c}_1(s)\cdot \mathbf{c}_2(s)$ reduces to 1, which can then be decrypted to recover the product $m_1 \cdot m_2$.
\end{itemize}

The proposed scheme is secure against an \textit{Indistinguishability Chosen Plaintext (IND-CPA)} attack.




\subsection{CKKS scheme}
The Cheon--Kim--Kim--Song (CKKS) scheme has been introduced in \cite{ckks} and is an asymmetric-key and leveled fully homomorphic scheme. 

Let $n$ be a power of 2, $\Delta >1$, and $\mathbb{C}$ be the set of complex numbers. Consider the isomorphism
\[
\sigma : \mathbb{C}[x]/(x^n+1) \longrightarrow \mathbb{C}^n
\]
defined as follows: let $\xi=\exp\Big(\frac{2 i \pi}{2n}\Big)$, then for every $f \in \mathbb{C}[x]/(x^n+1)$
\[
\sigma(f)=(f(\xi), f(\xi^3), \ldots, f(\xi^{2n-1})) \in \mathbb{C}^n.
\]
For every $\textbf{z}=(z_1,\ldots,z_{n/2}) \in \mathbb{C}^{n/2}$, let
\[
\pi(\textbf{z})=(z_1,\ldots,z_{n/2},\overline{z_{n/2}},\ldots,\overline{z_{1}}) \in \mathbb{C}^n,
\]
where $\overline{z_i}$ denotes the conjugate of $z_i$. To the vector $\Delta \cdot \pi(\textbf{z})$ we associate the vector
\[
\textbf{a}=(a_1, \ldots, a_n),
\]
with $a_i=\frac{\langle \textbf{z}, \textbf{v}_i \rangle}{\langle \textbf{v}_i, \textbf{v}_i \rangle}$ and $\textbf{v}_i$ the $i$-th row of the Vandermonde matrix
\[
A= \begin{pmatrix}
(\xi)^0 & (\xi)^1 & (\xi)^2 & \ldots & (\xi)^n \\
(\xi^3)^0 & (\xi^3)^1 & (\xi^3)^2 & \ldots & (\xi^3)^n \\
(\xi^5)^0 & (\xi^5)^1 & (\xi^5)^2 & \ldots & (\xi^5)^n \\
\vdots & \vdots & \vdots & \ddots & \vdots \\
(\xi^{2n-1})^0 & (\xi^{2n-1})^1 & (\xi^{2n-1})^2 & \ldots & (\xi^{2n-1})^n \\
\end{pmatrix}.
\]
Let $\tilde{\textbf{a}}=(\lceil a_1 \rfloor, \ldots, \lceil a_n \rfloor) \in \mathbb{Z}^n$. To the vector $\textbf{z} \in \mathbb{C}^{n/2}$ corresponds the polynomial
\[h=\sigma^{-1}(\tilde{\textbf{a}}) \in \mathbb{Z}[x]/(x^n+1),\]
where $\sigma^{-1}(\tilde{\textbf{a}})$ is the polynomial whose coefficient vector is $A^{-1}\tilde{\textbf{a}}$.

Let $g \in \mathbb{Z}[x]/(x^n+1)$, then $\pi^{-1}\Big(\sigma(\Delta^{-1} \cdot g)\Big) \in \mathbb{C}^{n/2}$.

Let $p$, $q_0$, $L$ be three positive integers and $q_{\ell}=p^{\ell} \cdot q_0$, with $0 \leq \ell \leq L$. Let $\mathcal{R}=\mathbb{Z}[x]/(x^n+1)$ and $\mathcal{R}_t=\mathbb{Z}_t[x]/(x^n+1)$. Let $\mathcal{X}(\sigma^2)$ be a Gaussian distribution with variance $\sigma^2$ over $\mathcal{R}_3$. Let $h$ be a positive integer, we denote by $HWT(h)$
the set of polynomials in $\mathcal{R}_3$ having Hamming weight equal to $h$, i.e., if $f=a_0+a_1x+\ldots+a_kx^k$ then
\[
|\{i : a_i \ne 0\}|=h.
\]
Let $\rho \in [0,1] \cap \mathbb{R}$, we denote by $\mathcal{Z}(\rho)$ the distribution that determines the coefficients of the polynomials in $\mathcal{R}_3$, with probability $\rho$ if $a_i \in \{-1, 1\}$ and probability $1 - \rho$ if $a_i =0$.

\begin{itemize}
    \item \textit{Keygen.} Let $\lambda$ be the security parameter. Let $n$ be a power of two, $h$ and $P$ integers, $\sigma$ a real number polynomial in $\lambda$.

    The algorithm chooses $s\in HWT(h)$ and $a\in \mathcal{R}_{q_L}$ (uniformly at random), $e$ from the distribution $\mathcal{X}(\sigma^2)$.

    The output is the secret key $\textbf{sk}=s$ and the public key $$\textbf{pk}=(-a\cdot s + e \pmod{q_L}, a) \in \mathcal{R}^2_{q_L}.$$

    \item \textit{Encrypt.} A message is a vector $\textbf{z} \in \mathbb{C}^{n/2}$. Let $m \in \mathcal{R}$ be the polynomial obtained from $\textbf{z}$ as described above. Let $\textbf{pk}=(\textbf{pk}[0], \textbf{pk}[1])$ be the public key. The algorithm chooses $v$ from the distribution $\mathcal{Z}(0.5)$ and $e_0, e_1$ from $\mathcal{X}(\sigma^2)$. The output is the ciphertext
    \[
    \textbf{c}=(v \cdot \textbf{pk}[0] + m+ e_0 \pmod{q_{\ell}}, v \cdot \textbf{pk}[1] + e_1 \pmod{q_{\ell}}).
    \]

    \item \textit{Decrypt.} Given the ciphertext $\textbf{c}=(\textbf{c}[0], \textbf{c}[1]) \in \mathcal{R}_{q_{\ell}}^2$ and the secret key $\textbf{sk}=s$, the output is given by
    \[
    \textbf{c}[0]+\textbf{c}[1] \cdot s \pmod{q_{\ell}},
    \]
    and corresponds to an approximate value of the message $m$. To the polynomial $m$ corresponds the vector $\pi^{-1} \Big( \sigma(\Delta^{-1} \cdot m) \Big) \in \mathbb{C}^{n/2}$.

    \item \textit{Eval.} Let $\textbf{c}_1, \textbf{c}_2 \in \mathcal{R}^2_{q_{\ell}}$ be the ciphertexts corresponding to the messages $m_1, m_2$, respectively.

    \textit{Add} $\rightarrow$ $\textbf{c}_{3}=\textbf{c}_1+\textbf{c}_2 \pmod{q_{\ell}}$ and we have $\textit{Encrypt}(m_1+m_2)=\textbf{c}_3$.

    \textit{Mult} $\rightarrow$ Let $\textbf{c}_1 = (b_1, a_1)$, $\textbf{c}_2 = (b_2, a_2) \in \mathcal{R}^2_{q_{\ell}}$. Let $(d_0, d_1, d_2) =
    (b_1 \cdot b_2, a_1 \cdot b_2+a_2\cdot b_1, a_1 \cdot a_2) \pmod{q_{\ell}}$. We introduce a new key called the \textit{evaluation key}: $$\textbf{evk}=(\textbf{evk}[0], \textbf{evk}[1]) \in \mathcal{R}_{P \cdot q_{L}}^2.$$ Chosen $a' \in \mathcal{R}_{P \cdot q_{L}}$ (uniformly at random), $e'$ from the distribution $\mathcal{X}(\sigma^2)$, the key $\textbf{evk}$ is given by
    \[
    \textbf{evk}[0]=-a' \cdot s + e'+ P \cdot s^2 \pmod{P \cdot q_{L}}, \quad \textbf{evk}[1]= a'.
    \]
    Then we define $\textbf{c}_{3}$ as the pair
    \[
    (d_0+ \lceil P^{-1} \cdot d_2 \cdot \textbf{evk}[0] \rfloor \pmod{q_{\ell}}, d_1+\lceil P^{-1} \cdot d_2 \cdot \textbf{evk}[1] \rfloor \pmod{q_{\ell}}).
    \]
    We have $\textit{Encrypt}(m_1 \cdot m_2)=\textbf{c}_{3}$

    \item \textit{Rescaling.} Let $\textbf{c} \in \mathcal{R}^2_{q_{\ell}}$ and $\ell'< \ell$, the algorithm outputs $$\textbf{c}'=\Big\lceil \frac{q_{\ell'}}{q_{\ell}} \textbf{c} \Big\rfloor \in \mathcal{R}^2_{q_{\ell'}}.$$ This algorithm allows to reduce the approximation error when performing multiplications between ciphertexts.
\end{itemize}

The scheme is secure against an \textit{Indistinguishability Chosen
Plaintext (IND-CPA)} attack.

\subsection{GSW scheme}

The Gentry--Sahai-Waters (GSW) scheme has been introduced in \cite{GSW} and is an asymmetric-key and leveled fully homomorphic scheme. 

Let $\mathrm{BitDecomp}$ be the operator which expands each coefficient of a vector into its binary representation. More precisely, if $\mathbf{a}=(a_1,\ldots,a_k) \in \mathbb{Z}_q^k$ and $\ell= \lfloor \log q \rfloor +1$, then $\mathrm{BitDecomp}(\mathrm{a})$ is the vector of dimension $k \ell$ obtained by replacing each entry $a_i$ with the list of its binary digits. If $A$ is a matrix, then $\mathrm{BitDecomp}(A)$ is obtained by applying $\mathrm{BitDecomp}$ to each row of $A$ separately.  

The operator $\mathrm{Flatten}$ is defined as the inverse of $\mathrm{BitDecomp}$. This operator groups blocks of $\ell$ coefficients and interprets each block as the binary expansion of an element of $\mathbb{Z}_q$. Thus, it reconstructs the original vector from its bit decomposition; moreover, it is also defined on inputs that are not necessarily binary. Again, for matrices the operation is applied row by row. 
For any vector $\mathbf{a}$, define
\[
\mathrm{Flatten}(\mathbf{a})
=
\mathrm{BitDecomp}\bigl(\mathrm{BitDecomp}^{-1}(\mathbf{a})\bigr).
\]

\begin{itemize}
    \item \textit{Keygen.} Let $\lambda$ be the security parameter. Choose $q,n$ and $m$ such that $m \in O(n \log q)$.  Let $\ell= \lfloor \log q \rfloor +1$ and $N =(n+1) \ell$. Let $\mathcal{X}$ be an error distribution over $\mathbb{Z}$.

    The algorithm samples uniformly at random $\mathbf{t}=(t_1,\ldots,t_n) \in \mathbb{Z}_q^n$ and outputs the secret key $$\mathbf{sk}=(1,-t_1,\ldots,-t_n) \in \mathbb{Z}_q^{n+1}.$$
    Put
    \[
\mathbf{v}
=
(1, 2, \dots, 2^{\ell-1},\;
 -t_1, -2t_1, \dots, -2^{\ell-1}t_1,\;
 \dots,\;
  -t_n,  -2t_n, \dots, -2^{\ell-1}t_n).
\]

    Then, it samples uniformly at random a matrix $B \in \mathbb{Z}_q^{m \times n}$ and from the distribution $\mathcal{X}^m$ an error vector $\mathbf{e} \in \mathbb{Z}^m$. Now, put $\mathbf{b}=B \cdot \mathbf{t}+ \mathbf{e}$. The public key \textbf{pk} is given by the $m \times (n+1)$-matrix $A$ whose first column is $\mathbf{b}$ and whose remaining $n$ columns are the columns of $B$. Note that $A \cdot \mathbf{sk}= \mathbf{e}$.

    \item \textit{Encrypt.} Let $\mu \in \mathbb{Z}_q$ be a message. Samples uniformly at random a matrix $R \in \mathbb{Z}_2^{N \times m}$ and outputs the ciphertext 
    $$
    C=\mathrm{Flatten}(\mu \cdot I_N + \mathrm{BitDecomp}(R \cdot A)) \in \mathbb{Z}_q^{N \times N},
    $$
    where $I_N$ is the identity matrix of order $N$. In particular, we have
    $$
    C \cdot \mathbf{v}= \mu \cdot \mathbf{v} + R \cdot A \cdot \mathbf{sk}= \mu \cdot \mathbf{v} + R \cdot \mathbf{e}. 
    $$

    \item \textit{Decrypt.} 
Among the entries of $\mathbf{v}$, consider one coefficient $v_i = 2^i$ such that
\[
v_i \in (q/4,\, q/2].
\]
Let $C_i$ denote the $i$-th row of the ciphertext matrix $C$. Then compute
\[
x_i = \langle C_i, \mathbf{v} \rangle,
\]
and recover the plaintext by 
\[
\mu = \left\lfloor \frac{x_i}{v_i} \right\rceil.
\]

\item \textit{Eval.} Let $C_1, C_2$ be the ciphertexts corresponding to the messages $\mu_1,\mu_2$, respectively.

\textit{Add} $\rightarrow$ Define
\[
C_3=\mathrm{Flatten}(C_1+C_2).
\]
Applying the decryption algorithm to $C_3$ recovers the message $\mu_1+\mu_2$.

\textit{Mult} $\rightarrow$  Define
\[
C_3=\mathrm{Flatten}(C_1 \cdot C_2).
\]

A crucial point in the GSW scheme is that homomorphic multiplication preserves correctness only if the resulting error term remains sufficiently small. Indeed, if
\[
C_1 \mathbf{v} = \mu_1 \mathbf{v} + e_1
\qquad\text{and}\qquad
C_2 \mathbf{v} = \mu_2 \mathbf{v} + e_2,
\]
then
\[
C_1 C_2 \mathbf{v}
= C_1(\mu_2 \mathbf{v} + e_2)
= \mu_2(\mu_1 \mathbf{v} + e_1) + C_1 e_2
=\mu_1 \mu_2 \mathbf{v} + \mu_2 e_1 + C_1 e_2.
\]
Hence, the product ciphertext still encrypts the message $\mu_1\mu_2$, but with a new error term
\[
e' = \mu_2 e_1 + C_1 e_2.
\]
Therefore, decryption remains correct only as long as $e'$ is sufficiently small.

\end{itemize}

The security of the scheme is based on the Learning With Errors (LWE) assumption. In particular, let $C$ be a ciphertext encrypting a message $\mu$. By construction, one has
\[
\mathrm{BitDecomp}^{-1}(C) = \mu G + RA,
\]
where $A$ is the public LWE matrix, $R$ is a random binary matrix, and $G$ is a fixed public matrix. Since $C$ is obtained by applying $\mathrm{BitDecomp}$ (or equivalently $\mathrm{Flatten}$) to $\mu G + RA$, it suffices to show that $\mu G + RA$ computationally hides the plaintext. This follows from the standard LWE argument: when $A$ is sampled together with an LWE secret and error, the resulting public key is computationally indistinguishable from uniform, and for $m \in O(n \log q)$ the distribution $(A,RA)$ is computationally indistinguishable from uniform as well. Therefore, ciphertexts do not reveal any information about $\mu$, and the scheme achieves semantic security under the LWE assumption. Moreover, by known reductions for LWE, this security can be related to the hardness of worst-case lattice problems such as $\mathrm{GapSVP}$ under appropriate parameter choices.

\subsection{TFHE/FHEW scheme} 

The FHEW scheme was introduced by Ducas and Micciancio in \cite{fhew}, while TFHE was introduced by Chillotti, Gama, Georgieva, and Izabach\`ene in \cite{tfhe}. These schemes are fully homomorphic and designed for fast bootstrapping of Boolean ciphertexts. In particular, FHEW showed that it is possible to bootstrap a Boolean gate in less than one second, while TFHE reformulated and improved this approach over the torus, significantly reducing the running time and the size of the bootstrapping key. The main idea is to evaluate arbitrary Boolean circuits gate by gate, refreshing the ciphertext after each non-linear operation by means of bootstrapping.

TFHE is defined over the torus
\[
\mathbb{T}=\mathbb{R}/\mathbb{Z},
\]
and uses three main types of ciphertexts:
\begin{itemize}
    \item \textit{TLWE} ciphertexts, which encrypt scalar messages over the torus;
    \item \textit{TRLWE} ciphertexts, which encrypt torus polynomials in the ring
    \[
    \mathcal{R}=\mathbb{T}[x]/(x^N+1),
    \]
    where $N$ is a power of $2$;
    \item \textit{TRGSW} ciphertexts, which are structured ciphertexts used to perform controlled multiplications and, in particular, blind rotations.
\end{itemize}

A TLWE ciphertext encrypting a message $\mu \in \mathbb{T}$ under a secret key $\textbf{s}\in \mathbb{Z}_2^n$ is a pair
\[
(\textbf{a},b)\in \mathbb{T}^n \times \mathbb{T},
\]
such that
\[
b=\langle \textbf{a},\textbf{s}\rangle+\mu+e,
\]
where $\textbf{a}$ is sampled uniformly at random and $e$ is an error term. Similarly, a TRLWE ciphertext encrypting a polynomial $\mu(x)\in \mathbb{T}[x]/(x^N+1)$ under a ring secret key $s(x)$ is a pair
\[
(a(x),b(x))
\]
such that
\[
b(x)=a(x)\cdot s(x)+\mu(x)+e(x),
\]
where $e(x)$ is a polynomial error. The TRGSW ciphertexts encrypt Boolean values or small integer polynomials and are used to homomorphically control operations on TRLWE ciphertexts.

In the FHEW/TFHE framework, the crucial operation is the bootstrapping procedure. In contrast with leveled schemes, where the multiplicative depth is fixed in advance, bootstrapping refreshes the noise of a ciphertext and therefore allows the evaluation of circuits of arbitrary depth. Moreover, in TFHE bootstrapping is \textit{programmable}: besides reducing the noise, it can also evaluate a function of the encrypted message, and this is the reason why Boolean gates can be implemented directly through bootstrapping.

The scheme can be described as follows.

\begin{itemize}
    \item \textit{Keygen.} Given a security parameter $\lambda$, the algorithm generates a secret key for TLWE encryption and the corresponding ring secret key for TRLWE/TRGSW operations. It also produces two public evaluation keys:
    \begin{itemize}
        \item a \textit{bootstrapping key}, consisting of TRGSW encryptions of the bits of the TLWE secret key;
        \item a \textit{key-switching key}, which is used to transform ciphertexts from one key or parameter set to another after bootstrapping.
    \end{itemize}

    \item \textit{Encrypt.} Let $m\in \{0,1\}$ be a 
message. The message is first encoded as a torus element, typically as one of two well-separated torus values, and then encrypted as a TLWE ciphertext
    \[
    c=(\textbf{a},b).
    \]
    The ciphertext contains the message masked by a small noise term.

    \item \textit{Decrypt.} Given a TLWE ciphertext $c=(\textbf{a},b)$ and the secret key $\textbf{s}$, the phase
    \[
    \varphi_{\textbf{s}}(c)=b-\langle \textbf{a},\textbf{s}\rangle
    \]
    is computed. Since this value is equal to the encoded message plus a small error, the plaintext is recovered by rounding $\varphi_{\textbf{s}}(c)$ to the nearest valid encoding.

    \item \textit{Eval.} Let $c_1$ and $c_2$ be two ciphertexts corresponding to the messages $m_1$ and $m_2$. 
    
    \textit{Add} $\rightarrow$ We have $c_1+c_2=Encrypt(m_1+m_2)$.
    
    \textit{Non-linear gates} $\rightarrow$ The evaluation of gates such as NAND, AND, OR, and MUX is achieved through \textit{gate bootstrapping}. In TFHE, this procedure consists of the following steps:
    \begin{itemize}
        \item initialization of an accumulator as a TRLWE encryption of a lookup table;
        \item \textit{blind rotation} of the accumulator, controlled by the encrypted phase of the input ciphertext and by the bootstrapping key;
        \item \textit{sample extraction}, which converts the resulting TRLWE ciphertext into a TLWE ciphertext;
        \item \textit{key switching}, which transforms the extracted ciphertext into the desired key and parameter set.
    \end{itemize}
    The result is a fresh TLWE ciphertext encrypting the output of the Boolean gate with reduced noise.
\end{itemize}

The main difference between FHEW and TFHE lies in the implementation of the bootstrapping step. FHEW is formulated over modular integer arithmetic and uses LWE, RLWE, and RGSW ciphertexts, while TFHE reformulates the scheme over the torus by means of TLWE, TRLWE, and TRGSW ciphertexts. Moreover, TFHE replaces the original internal-product based approach by an external product between a GSW-type ciphertext and an LWE-type ciphertext, which greatly improves the efficiency of blind rotation and reduces both the bootstrapping time and the size of the bootstrapping key.

The security of the FHEW/TFHE family is based on the hardness of the Learning With Errors problem and of its ring variants. More precisely, the basic ciphertexts rely on LWE/TLWE hardness, while the ring ciphertexts rely on Ring-LWE type assumptions. As in other bootstrapped FHE schemes, the publication of evaluation keys encrypting the secret key is typically justified through a circular-security assumption. Under suitable parameter choices, these schemes are considered secure against chosen-plaintext attacks, while the use of bootstrapping makes them fully homomorphic by allowing an unbounded number of homomorphic operations.

\subsection{DGHV scheme}

The DGHV scheme has been introduced by 
van Dijk, Gentry, Halevi and Vaikuntanathan in \cite{dghv}. The scheme is asymmetric-key and somewhat homomorphic.

\begin{itemize}
    \item \textit{Keygen.}  The construction depends on four parameters, each assumed to be polynomial in the security parameter $\lambda$: 
    \begin{itemize}
    \item $\gamma$ is the bit-length of the integers in the public key;
    \item $\eta$ is the bit-length of the secret key;
    \item $\rho$ is the bit-length of the noise;
    \item $\tau$ is the number of integers in the public key.
    \end{itemize}
    The parameters of the scheme are required to satisfy several asymptotic conditions. First, the noise parameter $\rho$ must grow faster than $\log \lambda$, so as to rule out exhaustive search attacks against the error term. Second, the secret-key length $\eta$ has to be sufficiently large compared with $\rho$, namely of order at least $\rho \cdot \Theta(\lambda \log^2 \lambda)$, in order to allow the evaluation of circuits of adequate depth. Furthermore, the public-key bit-length $\gamma$ must be super-quadratic in $\eta$ (up to logarithmic factors), in order to resist known lattice attacks against the approximate common divisor problem. Finally, the number $\tau$ of public-key elements must be taken large enough, namely at least $\gamma+\omega(\log \lambda)$, so that the leftover hash lemma can be applied in the reduction to the approximate-gcd assumption.

    In addition, one introduces an auxiliary noise parameter
\[
\rho'=\rho+\omega(\log \lambda).
\]
A convenient asymptotic choice of parameters is, for example,
\[
\rho=\lambda,\qquad \rho'=2\lambda,\qquad \eta=\tilde{O}(\lambda^2),\qquad \gamma=\tilde{O}(\lambda^5),\qquad \tau=\gamma+\lambda.
\]
With this choice, the overall complexity of the scheme is $\tilde{O}(\lambda^{10})$.

For a fixed odd positive integer $p$ of bit-length $\eta$, consider the following probability distribution over $\gamma$-bit integers:
\[
D_{\gamma,\rho}(p)=
\left\{
\begin{array}{l}
\text{sample } q \text{ uniformly from } \mathbb{Z}\cap [0,2^\gamma/p),\\[1mm]
\text{sample } r \text{ uniformly from } \mathbb{Z}\cap (-2^\rho,2^\rho),\\[1mm]
\text{output } x=pq+r.
\end{array}
\right.
\]

Sample the secret key as an odd $\eta$-bit integer
    \[
    \mathbf{sk}=p \in (2\mathbb{Z}+1)\cap [2^{\eta-1},2^\eta).
    \]
    Then generate public-key elements $ x_i, i=0,\ldots,\tau,$ from the distribution $D_{\gamma,\rho}(p)$. Reorder them so that $x_0$ is the largest one. If $x_0$ is not odd or if its residue modulo $p$ is not even, restart the procedure. The public key is
    \[
    \mathbf{pk}=( x_0,x_1,\ldots,x_\tau).
    \]

    \item \textit{Encrypt.} Let $m \in \mathbb{Z}_2$ be a message. Choose uniformly at random a subset
    \[
    S\subseteq \{1,2,\ldots,\tau\}
    \]
    and sample
    \[
    r \in (-2^{\rho'},2^{\rho'}).
    \]
    The ciphertext is
    \[
    \mathbf{c}= m+2r+2\sum_{i\in S}x_i \pmod{x_0}.
    \]
    Note that, by the definition of the public key elements $x_i$, the ciphertext $\mathbf{c}$ can be represented in the form:  $$\mathbf{c} = m + 2\beta + \alpha p.$$

    \item \textit{Decrypt.} Recover the message by computing
    \[
    (c \bmod p)\bmod 2.
    \]

    \item \textit{Eval.} Let $\textbf{c}_1= m_1+2r_1+2\sum_{i\in S_1}x_i \pmod{x_0}$ and $\textbf{c}_2= m_2+2r_2+2\sum_{i\in S_2}x_i \pmod{x_0}$ be the ciphertexts corresponding to the messages $m_1$ and $m_2$, respectively.

\textit{Add} $\rightarrow$ $\textbf{c}_3=\textbf{c}_1+\textbf{c}_2$ and we have $\textbf{c}_3=Encrypt(m_1+m_2)$.

\textit{Mult} $\rightarrow$ $\textbf{c}_3=\textbf{c}_1\cdot\textbf{c}_2$ and we have $Encrypt(\textbf{c}_3)=m_1 \cdot m_2$.
\end{itemize}

The security of the DGHV scheme relies on the hardness of the approximate greatest common divisor problem. Indeed, the public-key integers are noisy multiples of a hidden odd integer \(p\), which forms the secret key. Thus, breaking the scheme amounts to recovering \(p\) from approximate multiples, a problem believed to be hard for suitable parameter choices.

\section{Conclusion}

This survey has reviewed homomorphic encryption schemes arising from two broad viewpoints: constructions based on coding theory and constructions based on polynomial. Although these two families are often presented separately, the comparison developed below shows that they are strongly connected at the algebraic level, while differing significantly in their security foundations, efficiency, and current practical relevance.

Table~\ref{tab:comparison-he-schemes} summarizes the main features of the schemes discussed in this paper. The comparison is organized according to the following criteria: scheme type (symmetric or asymmetric; PHE/SHE/FHE), main hardness assumption or algebraic structure, supported multiplicative depth, current security status, and practical relevance.

\begin{table}[htbp]
\centering
\scriptsize
\setlength{\tabcolsep}{3pt}
\renewcommand{\arraystretch}{1.15}
\begin{tabular}{|p{2.3cm}|p{2.3cm}|p{2.6cm}|p{2.2cm}|p{2.2cm}|p{2.4cm}|}
\hline
\textbf{Name} & \textbf{Scheme type} & \textbf{Main assumption / structure} & \textbf{Multiplicative depth} & \textbf{Security status} & \textbf{Practical relevance} \\
\hline
Armknecht et al. & Symmetric; SHE & Evaluation codes; DSCP-type security intuition & Bounded by parameter $\mu$ & Non-standard security basis; limited modern validation & Primarily conceptual \\
\hline
Challa--Gunta (vector version) & Symmetric; FHE & Reed--Muller codes; hiding positions of codeword symbols & Unbounded & Non-standard security claim; limited supporting evidence & No practical deployment \\
\hline
Challa--Gunta (matrix version) & Symmetric; FHE & Reed--Muller codes with secret error positions and permutation & Unbounded & Non-standard security claim; limited supporting evidence & No practical deployment \\
\hline
Bogdanov--Lee & Asymmetric; SHE & Modified Reed--Solomon structure & Bounded & Broken & None \\
\hline
Aguilar--Melchor et al. & Symmetric; SHE & Random rank-metric ideal codes; IRSD & One ciphertext multiplication (plus additions / plaintext absorptions) & Security depends on publication bounds; bootstrap proposal insecure & Mainly theoretical \\
\hline
Dasgupta--Pal & Symmetric; SHE; bootstrappable & Polynomial representation over integers; AGCD & Bounded before refresh & Mainly theoretical & Conceptual link to integer-based HE \\
\hline
Dowerah--Krishnaswamy & Symmetric; SHE & Multivariate polynomial evaluation with hidden ideal structure & Bounded & Claimed IND--CPA; non-standard framework & Mainly theoretical \\
\hline
BFV & Asymmetric; leveled FHE & RLWE / ring arithmetic & Bounded by chosen level & Standard and widely studied & Practical for exact arithmetic \\
\hline
BGV & Asymmetric; leveled FHE & LWE/RLWE with modulus switching & Bounded by chosen level & Standard and widely studied & Practical and influential \\
\hline
CKKS & Asymmetric; leveled FHE (approximate) & RLWE / approximate arithmetic & Bounded by chosen level & Standard and widely studied & Highly relevant in applications involving real/complex data \\
\hline
GSW & Asymmetric; leveled FHE & LWE & Bounded before bootstrapping & Standard and foundational & Mainly foundational; important in later bootstrapping frameworks \\
\hline
FHEW / TFHE & Asymmetric; FHE via bootstrapping & LWE / RLWE / torus variants & Effectively unbounded through gate bootstrapping & Standard and widely studied & Highly relevant for Boolean circuits and fast bootstrapping \\
\hline
DGHV & Asymmetric; SHE; bootstrappable to FHE & Approximate GCD & Bounded before bootstrapping & Foundational but less practical than lattice-based schemes & Mainly historical and conceptual \\
\hline
\end{tabular}
\caption{Comparative overview of the homomorphic encryption schemes discussed in this survey.}
\label{tab:comparison-he-schemes}
\end{table}

The table makes evident a major divide in the literature. On the one hand, many code-based proposals are algebraically elegant and conceptually important, but often rely on less standardized security assumptions, and several of them remain primarily theoretical or have been shown to be insecure. On the other hand, lattice- and ring-based schemes such as BFV, BGV, CKKS, GSW, and TFHE/FHEW are built on assumptions that are much more established in modern post-quantum cryptography, and they have also led to practical implementations and software libraries.

The distinction between the two families considered in this survey should not be interpreted too rigidly. In fact, several codes appearing in Section \ref{coding} are themselves defined through polynomial evaluation. Reed--Muller codes arise from the evaluation of multivariate polynomials of bounded degree, Reed--Solomon codes arise from the evaluation of univariate polynomials, and some rank-metric constructions are naturally related to polynomial or ideal-module representations. From this viewpoint, coding-theoretic schemes and polynomial-based schemes share a common algebraic core: both encode information into structured algebraic objects and exploit the compatibility of this structure with addition and multiplication.

The difference lies mainly in the language in which correctness and security are formulated. In coding-theoretic proposals, the ciphertext is typically interpreted as a codeword or a noisy codeword, and decryption is linked to interpolation or decoding from hidden error positions or hidden supports. In polynomial- and ring-based schemes, the ciphertext is more naturally viewed as a polynomial or a tuple of ring elements with embedded noise, and decryption relies on recovering the message from a controlled noisy algebraic relation. Thus, the contrast is not between ``codes'' and ``polynomials'' as disjoint objects, but rather between two different cryptographic paradigms: one expressed primarily in terms of decoding problems, and the other in terms of noisy arithmetic over structured rings.

This observation also clarifies why some schemes occupy an intermediate position. For example, the Dasgupta--Pal and DGHV schemes are both based on arithmetic over the integers with hidden approximate divisibility relations, while rank-metric and ideal-code proposals combine coding-theoretic language with algebraic operations that resemble polynomial-ring methods. Likewise, lattice-based schemes such as BFV, BGV, CKKS, GSW, and TFHE/FHEW may be viewed as exploiting highly structured codes over rings, although their modern security analysis is framed through LWE, RLWE, or related assumptions rather than classical decoding problems.

Therefore, the two families surveyed in this paper should be regarded not as disconnected lines of research, but as points on a common spectrum of algebraic homomorphic encryption. The unifying principle is the search for algebraic representations that simultaneously support: (i) efficient encryption and decryption; (ii) homomorphic addition and multiplication; and (iii) a hard underlying inversion problem for an adversary.

\section{Open problems}

The schemes reviewed in this survey illustrate the evolution of homomorphic encryption from early algebraic and code-based proposals to the modern dominance of lattice- and ring-based constructions. A first general lesson is that the mere existence of an appealing algebraic structure is not sufficient: to obtain a convincing homomorphic encryption scheme, one also needs a robust and well-understood hardness assumption, together with a precise control of noise growth under evaluation. This is precisely where many early or non-standard constructions become fragile.

A second lesson is that practical relevance currently belongs mainly to lattice-based schemes. BFV and BGV provide exact arithmetic over finite rings, CKKS supports approximate arithmetic for numerical computation, and TFHE/FHEW offer extremely efficient bootstrapping for Boolean circuits. By contrast, most code-based constructions surveyed here remain mainly of theoretical interest, either because their security assumptions are less mature, or because efficient and secure bootstrapping mechanisms are still missing.

At the same time, code-based homomorphic encryption remains an important research direction. It is closely related to post-quantum cryptography, it raises natural decoding problems of independent mathematical interest, and it may still offer alternative design paradigms that differ substantially from the lattice-based mainstream. In this sense, the negative results and attacks discussed in the paper should not be viewed only as limitations, but also as guidance for future design principles.

Several open problems emerge naturally from the comparison developed in this survey.

\begin{itemize}
    \item \textbf{Standard-assumption code-based homomorphic encryption.} One of the main challenges is to construct code-based homomorphic encryption schemes whose security is based on assumptions that are as clear and as broadly accepted as LWE or RLWE.

    \item \textbf{Secure and efficient bootstrapping outside the lattice setting.} A decisive step toward practically relevant code-based FHE would be the design of a bootstrapping procedure that is both efficient and supported by a convincing security analysis.

    \item \textbf{Sharper security analyses for non-standard proposals.} Several schemes in the literature rely on ad hoc hiding arguments, secret positions, or algebraic obfuscation effects. A systematic reformulation of such claims in modern indistinguishability-based terms would be highly desirable.

    \item \textbf{Bridging evaluation-code methods and ring-based techniques.} Since many coding-theoretic objects are polynomial-evaluation structures, it is natural to ask whether the gap between code-based and ring-based homomorphic encryption can be narrowed through hybrid constructions.

    \item \textbf{Complexity and practicality.} Another important problem is to understand whether alternative algebraic frameworks can compete in practice with BFV, BGV, CKKS, or TFHE, either for exact arithmetic, approximate arithmetic, or Boolean computation.

    \item \textbf{Systematic comparison criteria.} The literature would also benefit from a more standardized methodology for comparing homomorphic schemes, taking into account not only asymptotic properties but also concrete security, ciphertext expansion, key size, evaluation cost, and applicability to real workloads.
\end{itemize}

In conclusion, the landscape of homomorphic encryption is best understood as an interaction between algebraic structure, noise management, and computational hardness. Coding theory and polynomial/ring methods should be viewed as complementary perspectives within this broader picture. While lattice-based schemes currently dominate both theory and practice, the algebraic ideas originating from coding theory continue to offer a rich source of questions, techniques, and possible future breakthroughs.

\section*{Acknowledgments}
This work was funded by the SERICS project (PE00000014) under the MUR
National Recovery and Resilience Plan funded by the European Union-NextGenerationEU.

 \noindent The author would like to thank the anonymous referee for the valuable comments and suggestions that helped improve the paper.

\end{document}